\documentstyle[preprint,epsf,aps,eqsecnum]{revtex}
\tightenlines
\newcommand{\dslash}{\partial\hspace{-.09in}/}
\newcommand{\kslash}{k\hspace{-.09in}/}

\begin{document}

\title{Neutrino oscillations in a model with a source and detector}

\author{Ken Kiers\footnote{Email: kiers@bnl.gov\vspace{-.15in}}}
\address{Department of Physics\\ Brookhaven National Laboratory,
Upton, NY 11973-5000, USA}

\author{Nathan Weiss\footnote{Email: weissn@post.tau.ac.il}}
\address{School of Physics and Astronomy, Tel Aviv University,\\
Tel Aviv, Israel\\Department of Particle Physics, Weizmann Institute,\\
Rehovot, Israel\\ Department of Physics and Astronomy,
University of British Columbia\\Vancouver, B.C., V6T 1Z1  Canada}
 
\maketitle
\begin{abstract}

We study the oscillations of neutrinos in a model in which the neutrino
is coupled to a localized, idealized source and detector. By varying the
spatial and temporal resolution of the source and detector we are able
to model the full range of source and detector types ranging from 
coherent to incoherent.  We find that this approach is useful in 
understanding the interface between the Quantum Mechanical nature
of neutrino oscillations on the one hand and the production and 
detection systems on the other hand. This method can easily be extended
to study the oscillations of other particles such as the neutral $K$
and $B$ mesons. We find that this approach gives a reliable way
to treat the various ambiguities which arise when one examines the
oscillations from a wave packet point of view. We demonstrate that
the conventional oscillation formula is correct in the relativistic limit
and that several recent claims of an extra factor of 2 in the oscillation
length are incorrect. We also demonstrate {\em explicitly} 
that the oscillations
of neutrinos which have separated spatially may be ``revived'' by a long
coherent measurement.

\end{abstract}

\pagebreak

\section{Introduction}
\label{sec:intro}

The flavour oscillations of particles are a fascinating
demonstration of quantum mechanics in the macroscopic world.
Flavour oscillations can generically occur when the states
which are produced and detected in a given experiment are
superpositions of two or more eigenstates which have different masses.  
The oscillations of $K$ and $B$ mesons have been observed 
experimentally~\cite{kosc}
and have been used to place stringent constraints on  
physics beyond the Standard Model.
If neutrinos are massive, they too may oscillate, and this
could lead to the resolution of the well-known solar neutrino 
problem~\cite{pontecorvo,wolfenstein,miksm}.
More recently, the discussion of particle oscillations has
been extended to include supersymmetric particles in 
supersymmetric extensions of the Standard Model~\cite{feng}.  

The phenomenon of particle oscillations has been studied extensively
and is generally thought to be very well understood.  There nevertheless
remain several subtle issues which continue to cause some confusion.
The key to a complete understanding of any such issue lays
in treating correctly the necessary interplay
between the ``classical'' and ``quantum'' natures of the particles
which are interfering to produce the oscillations.  Thus, for example,
the interference effect itself is purely ``quantum'' in nature (it
requires that the particles be described by waves), and yet 
the resulting oscillations in space are only observable if the particles
are sufficiently localized in space~\cite{kayser}.  This example
highlights the fact that any discussion of particle oscillations implicitly
assumes that the mass eigenstates which are interfering to produce the 
oscillations are described by some sort of wave packets.

Despite the success of the wave packet approach in clarifying many
aspects of the phenomenon of particle 
oscillations~\cite{kayser,nuss1,kimpacket1,kimpevsner}, the approach is 
not without its own difficulties.  The results of a
given calculation will depend, for example, on the details of 
the initial mass eigenstate wave packets (including their shape, 
spectrum and relative normalization).
One particularly difficult problem which arises  
is the conversion of 
the final time-evolved wave packets into an experimentally observable 
quantity:  since it is generally the {\bf flux} of particles 
which is measured in an experiment, one is required to calculate
a current density rather than a probability 
density.\footnote{Wave packet calculations lead naturally to expressions
for the probability density, which are appropriately integrated over
space, not time.  For oscillations in space one wants a quantity which
is appropriately integrated over time, i.e., a current.}
The difference between a current density and a probability
density, at least naively,
involves a factor of the velocity $v$, which is very significant
if the mass eigenstates have quite different masses.  
Thus, if one would calculate the probability density
at the detector and
integrate it over time, the resulting expression would have factors
of $1/v$ pre-multiplying the various terms, leading to an
enhancement of the terms corresponding to heavier mass eigenstates.
In the case of neutrinos, as was noted 
in~\cite{kimpacket1,kimpevsner,kimpacket1a},
this would skew the usual
oscillation formula quite dramatically if one of the mass
eigenstate neutrinos was non-relativistic.  Efforts to
construct an appropriate current density which retains the necessary
wave packet features have had mixed success.  A calculation in
the kaon case appears to give reliable results~\cite{ancochea},
but it can be shown that unphysical effects arise if one
attempts to define a suitable current when the mass eigenstates 
have very different masses~\cite{thesis}.
 
There is another very striking apparent ``ambiguity'' which arises if one
does not treat the delicate interplay between the classical and
quantum natures of the particles correctly.  In this case
the ``ambiguity'' leads to an alleged error by a factor of ``2'' 
in the calculation of the
oscillation length~\cite{factwo}.
In order to understand the source of the ambiguity, we
follow the discussion given by Lipkin~\cite{lipkin}.
Suppose we consider the oscillations in time of a system for which
the initial and final states are not eigenstates of the free Hamiltonian.
The phase of the interference term will then be
given by $\phi(t)$$=$$(E_1-E_2)t$, where $E_i$$\equiv$$(p^2+m_i^2)^{1/2}$.
Detectors do not measure oscillations directly as a function of time, however,
so one needs somehow to convert this expression into an oscillation
in terms of space.  We may then,
as is conventionally done, set $x$$=$$vt$, 
with $v$$=$$2p/(E_1+E_2)$ representing a sort
of average velocity.  We then obtain the following phase in terms of $x$,
\equation
	\phi_{\rm conv}(x) = \frac{(m_1^2-m_2^2)x}{2p} .
\label{eq:phaseconv}
\endequation
This is the conventional (and correct) result for the phase difference.
Let us now attempt to incorporate a classical aspect of the problem
and argue that since the two mass eigenstates
travel at different speeds, they will arrive at the detector
at {\em different} times $t_1$ and $t_2$, related by
$x$$=$$pt_1/E_1$$=$$pt_2/E_2$.  Taking the phase of the interference
term to be $\phi(t_1,t_2)$$=$$E_1t_1-E_2t_2$, we then obtain
\equation
	\phi_{\rm new}(x) = \frac{(m_1^2-m_2^2)x}{p} ,
\label{eq:phasenew}
\endequation
which differs from the conventional phase difference, 
Eq.~(\ref{eq:phaseconv}), by a factor of two.  This
result, were it correct, would indeed be rather remarkable.

The first resolution of this ambiguity was
given by Lipkin~\cite{lipkin} (see also
Ref.~\cite{grossman}) who
argued, on physical grounds, that the energies (rather than the momenta) of 
the two mass eigenstates should be set equal.  In this case the 
oscillations are described in terms of distances directly (since
it is the momenta of the two eigenstates which differ) and the ``correct''
oscillation formula is obtained. 
Lowe {\em et al}~\cite{lowe} and Kayser~\cite{kayser2}
have extended this discussion and have argued that 
the key to avoiding ambiguities is to ensure that one
evaluates the wave functions of the mass eigenstates
at precisely the same space-time point.  That is,
even though classically the mass eigenstates will arrive
at the detector at different times, quantum mechanically the
wave functions corresponding
to different space-time points cannot interfere. 
Indeed  the analysis leading to the 
expression in Eq.~(\ref{eq:phasenew}) involves interfering
the wave functions for the two mass eigenstates at the same 
position but {\em different} times and hence gives the incorrect result.
This issue is, in fact, quite subtle.  For example
a long coherent measurement in time may be used to ``revive'' particle
oscillations even after the mass eigenstate wave packets have completely
separated spatially~\cite{knw}.\footnote{This behaviour is analogous
to what happens when a high $Q$ oscillator gets hit by two
successive pulses.  The first pulse sets the oscillator in motion,
causing it to oscillate for a time determined by its $Q$ value.
If the oscillator is still oscillating when the second pulse arrives,
the resulting oscillations will exhibit an interference pattern.}
Thus, wave packets arriving at the ``classically'' separated
times $t_1$ and $t_2$ -- and having negligible overlap in their wave packets --
may still interfere to give rise to oscillations.  There is then
some sense in which wave functions corresponding to different space-time
points may interfere.

In light of the issues presented above, it is our view that a proper 
treatment of the Quantum -- Classical interface of particle oscillations 
should incorportate the source and the detector as key components of the
system.
In this paper we present a simple model for a particle
source-detector system which addresses many of the above
issues in a very natural and self-consistent way.  We shall,
for concreteness, consider the case of neutrino oscillations, but 
our approach could easily be adapted to other situations.  The
source and detector will be modeled by simple harmonic 
oscillators which are de-excited or excited by emitting
or absorbing neutrinos of a given flavour.  
(Two-level ``fermionic'' source -- detector systems could
also be considered.) Having defined
the model, it will be straightforward to calculate the 
oscillation probability as a function of the distance
between the source and detector.  The resulting 
expressions will be found to exhibit all of the known
``wave packet'' characteristics in the relativistic
limit, but will also give useful insight into cases
in which one or more of the mass eigenstates is non-relativistic.
In particular, we will find no evidence for the  
enhancement of non-relativistic neutrinos which can occur
in conventional wave packet analyses.  
Including the source explicitly in the calculation gives
the added benefit that  the 
characteristics of the initial wave packets corresponding to the
various mass eigenstates are completely determined
by the characteristics of the source and need not be put in by hand.
Our approach is similar in spirit to the calculations in
Refs.~\cite{gkll,grimus}, but is more transparent due to the
simplified model which we consider. (See also Ref.~\cite{rich} for 
a similar calculation performed within the context of elementary
quantum mechanics.)  One advantage of our simplified 
approach is that the dependence on the time-resolution of the detector is 
very clear.  This will allow us to verify explicitly that a long coherent
measurement in time may be used to revive the oscillations of 
particles whose wave packets have separated spatially.\footnote{A
recent paper has also demonstrated this effect explicitly~\cite{gkl97}.}
We shall also settle the issue of the ``factor of 2'' (hopefully)
once and for all.
 
We begin in the next section by analyzing a simple model
in which the neutrino is described by
a complex scalar field.  This field is coupled to two 
localized simple harmonic
oscillators, representing the source and detector.  Modeling the neutrino
by a complex scalar field
allows for a simpler and more complete evaluation of physical
quantities than if a spinor field is used.
In Sec.~\ref{asingleneut} we study the case of
a single neutrino species coupled to the source and detector.  This
allows for a careful analysis of the efficiency of our
system at producing and detecting neutrinos of different masses.
Although no oscillations are possible in this case, this calculation
will be essential in interpreting the results when neutrino
oscillations are present.
In Sec.~\ref{sevneut} we couple several neutrino fields
to the source and detector.  This gives rise in a natural
way to oscillations (as a function of the distance between the
source and detector) in the probability for the 
source to decay and the detector to be excited.  These are
of course ``neutrino oscillations.''
Sec.~\ref{sec:nonrelcase} contains a brief analysis of the non-relativistic
case.  We then extend our analysis in Sec.~\ref{sec3.3}
to a more realistic model in which 
the neutrinos are described by Dirac spinor fields.  These results
are compared to the ones with a complex scalar field.  We conclude
in Sec.~\ref{sec3.4} with a summary and discussion of our results.

\section{A Model for a Neutrino Source and Detector}
\label{sec3.2}

The idea of using an idealized detector to clarify physically
measurable quantities in Quantum Field Theory has been used extensively
in the analysis of Quantum Fields in non-inertial frames and in 
gravitational backgrounds~\cite{unruh2}.
In our idealized model, we have chosen to couple the neutrino field to
two harmonic oscillators, one representing a neutrino
``source,'' and the other representing a neutrino ``detector.''
The neutrinos are first taken to be complex scalar fields which
simplifies the calculations considerably.\footnote{
The main drawback of this approach is that it ignores the
neutrino's spin and the characteristic $V-A$ nature
of neutrino interactions.}

The physical picture which we have in mind is the following:
we imagine our ``source'' and ``detector'' to be microscopic
on the scale of some macroscopic ``bulk'' source and detector, but
to also be very massive compared to the energy of the exchanged neutrino
(so that the dynamical degrees of freedom of the source and detector
may be ignored).  Thus, for example, the
source (detector) could represent some nucleus inside a bulk sample
which undergoes beta decay (inverse beta decay).  
The spatial ``widths'' of the source and detector in our calculation are
then widths appropriate to, say, nuclear or atomic dimensions.  In principle,
the oscillation probability which we calculate here should subsequently
be averaged incoherently over the physical dimensions of the macroscopic
source and detector, although we do not perform this average.  If 
the size of the macroscopic source and detector are much smaller
than the neutrino oscillation length (which they need to be in
order to observe oscillations), then this averaging would have
only a small effect.

The interactions at the
source and detector will be made explicitly time dependent so that
they may be turned ``on'' and ``off.''  
This is in keeping with our physical picture.  
In general a real (microscopic) source or detector will be in an environment
which is ``noisy,'' so that the coherent emission or absorption
of a neutrino gets cut off after some time due to the interactions
of the source or detector with its surrounding 
environment~\cite{kimpevsner,knw}.
The amount of time which the model source or detector spend being ``on''
is then related to the ``coherence time'' of the physical source
or detector.\footnote{
The explicit turning
on and off of the source and detector violates energy
conservation microscopically, but that is natural since
the interactions of the source and detector with
their respective environments involve the exchange of energy.  If we choose 
to look at the source or detector in isolation, this exchange of
energy appears as energy non-conservation.}

Our calculation proceeds as follows.
We first write a Lagrangian which couples the source and detector
to the neutrino field.  In the initial state of the system, the source is
in its first excited state (ready to emit a neutrino) and the detector is in
its ground state.  We then calculate the probability that
at some time far in the future the source is found to be in
its ground state and the detector in its first excited state.
The model will be constructed in such a way that this interaction
will correspond to exactly one neutrino being exchanged between 
the source and detector (to first non-vanishing order in
perturbation theory.)  In this approach, then, 
the neutrinos themselves are not observed, but are simply the exchange
particles in the source-detector interaction.  

\subsection{A Single Species of Neutrino}
\label{asingleneut}

To describe our model, we begin with a single complex scalar field
$\phi(x)$ and two oscillators $q_1(t)$ and $q_2(t)$ describing the
source and detector, respectively.  The action for our model
is given by
\equation
	S = \int d^4x \left({\cal L}_{\phi}^0 + {\cal L}_{\rm int}\right)
		+ \int dt L_q^0,
	\label{toyaction}
\endequation
where
\begin{eqnarray}
	{\cal L}_{\phi}^0 & = & 
		-\phi^{\dagger}(x)\left(\Box + m^2\right)\phi(x), \\
	L_q^0 & = & \dot{q}_1^{\dagger}(t)\dot{q}_1(t) 
		- \Omega_1^2q_1^{\dagger}(t)q_1(t)
		+\dot{q}_2^{\dagger}(t)\dot{q}_2(t) 
		- \Omega_2^2q_2^{\dagger}(t)q_2(t),\\
	{\cal L}_{\rm int} & = & 
	        -\epsilon_1(t)\left(\phi^{\dagger}(x)q_1(t)h_1({\bf x})+
			\phi(x)q_1^{\dagger}(t)h_1^*({\bf x})\right)
		\nonumber \\
		& & 
		-\epsilon_2(t)\left(\phi^{\dagger}(x)q_2(t)h_2({\bf x})+
                        \phi(x)q_2^{\dagger}(t)h_2^*({\bf x})\right).
\end{eqnarray}
The functions $\epsilon_i(t)$
are explicit functions of time which allow us to ``turn on'' and ``turn 
off'' the interactions, and the functions $h_1({\bf x})$ 
($h_2({\bf x})$) are
smooth functions of ${\bf x}$ which vanish outside the source
(detector).  

We quantize the {\bf free} fields in the usual way, requiring
\begin{eqnarray}
	\left[\phi({\bf x},t),\pi({\bf y},t)\right] & = & 
		i \delta^3({\bf x}-{\bf y}) ,\\
	\left[q_i(t),p_i(t)\right] & = & i .
\end{eqnarray}
All other commutators are taken to vanish.  The field operators
may then be expressed in terms of creation and annihilation operators
as follows
\begin{eqnarray}
	\phi(x) & = &  \int d\tilde{k}\left( a(k) e^{-ik\cdot x}
		+b^{\dagger}(k) e^{ik\cdot x}\right) , 
	\label{phimode}	\\
	q_i(t) & = & \frac{1}{2\Omega_i}\left( A_i e^{-i\Omega_i t}
		+B_i^{\dagger} e^{i\Omega_i t}\right),
	\label{pmode}
\end{eqnarray}
where
\equation
	d\tilde{k} \equiv \frac{d^3k}{(2\pi)^3 2E}
\endequation
and where the annihilation and creation operators satisfy the
commutation relations
\begin{eqnarray}
	\left[a(k),a^{\dagger}(k^{\prime})\right] & = &
		\left[b(k),b^{\dagger}(k^{\prime})\right] =
		(2\pi)^3 2 E \delta^3 ({\bf k}-{\bf k}^{\prime}) ,
	\label{acom} \\
	\left[A_i,A_i^{\dagger}\right] & = &
		\left[B_i,B_i^{\dagger}\right] = 2 \Omega_i .
	\label{acapcom}
\end{eqnarray}
We interpret $a^{\dagger}(k)$ and $a(k)$ in the usual way
as the operators which create and annihilate, respectively, a neutrino
state with four-momentum $k$.  $b^{\dagger}(k)$ and $b(k)$ act similarly
with respect to the anti-neutrino states.  The operators $A_i^{\dagger}$
and $A_i$ and $B_i^{\dagger}$ and $B_i$ 
interpolate between the energy levels of the 
harmonic oscillators.\footnote{
Note that we have allowed the $q_i$ to be complex.
Had we not done this, the source and detector would have exchanged both
neutrinos and anti-neutrinos.}

We take as our initial state
\equation
	|s,-\infty\rangle = |0;1;0\rangle 
		\equiv |0\rangle_{\phi}\otimes 
		|1\rangle_1 \otimes|0\rangle_2
\endequation
in which
\equation
	|1\rangle_i \equiv A_i^{\dagger}|0\rangle_i
	\label{statedef}
\endequation
represents the first excited state of the oscillator $i$ and in
which $|0\rangle_{\phi}$ is the neutrino vacuum state.
We wish to calculate the amplitude for the process in which
the source de-excites to its ground state and the detector is
excited to its first excited state.  That is,
\equation
	{\cal A} \equiv \langle 0;0;1|s,\infty\rangle 
	= \langle 0;0;1|T {\rm exp}\left[
                -i\int_{-\infty}^{\infty} H^S (t^{\prime})dt^{\prime}
                        \right] |s,-\infty\rangle ,
	\label{oscamp}
\endequation
in which $H^S$ represents the Hamiltonian in the Schr\"{o}dinger
picture.
The modulus squared of this amplitude is the probability for
the transition to take place.

We shall assume the couplings in the interaction Hamiltonian
to be sufficiently small that 
the amplitude in Eq.~(\ref{oscamp}) is always much less than
unity.  This is of course always the case in the real-world
situation which we are attempting to model -- neutrino
interactions are so weak that perturbation theory is always
valid.  It is then straightforward to evaluate (\ref{oscamp})
using standard techniques to obtain to leading order and 
up to an over-all unobservable phase,
\equation
	{\cal A} = -\frac{1}{2}\langle 0;0;1|T
		\left[\int_{-\infty}^{\infty}dt^{\prime}dt^{\prime\prime}
		H_{\rm int}^H(t^{\prime})H_{\rm int}^H(t^{\prime\prime})
		\right] |0;1;0\rangle ,
	\label{adefined}
\endequation
where $H_{\rm int}^H(t)$ refers to the interaction Hamiltonian
evaluated in terms of the free fields in the Heisenberg picture
at time $t$.
The above expression may be evaluated explicitly in terms
of neutrino propagators~\cite{gkll,grimus} for 
arbitrary turn-on/off functions $\epsilon_i(t)$.  We find
it simpler, however, to ``design'' the turn on/off functions so that
the source and detector
are never on at the same time, and, furthermore,
so that the source  always turns on first and only then the
detector.  (This avoids the unphysical situation in which the 
detector emits an anti-neutrino which is subsequently absorbed by
the source.  The amplitude for this process would in any case be
very small since it violates energy conservation.)
Under this assumption only one of the time-orderings in the propagator
gets picked up and ${\cal A}$ may be evaluated 
using Eqs.~(\ref{phimode}), (\ref{pmode}), (\ref{acom}), (\ref{acapcom})
and (\ref{statedef}) to obtain
\begin{eqnarray}
	{\cal A} & = & - \langle 0;0;1|\int dt^{\prime}
		dt^{\prime\prime} d^3x^{\prime}
                d^3x^{\prime\prime} \epsilon_1(t^{\prime})
                \epsilon_2(t^{\prime\prime}) \nonumber \\
		& & \;\;\times
		\phi(x^{\prime\prime})q_2^{\dagger}(t^{\prime\prime})
		h_2^*({\bf x}^{\prime\prime})
		\phi^{\dagger}(x^{\prime})
		q_1(t^{\prime})h_1({\bf x}^{\prime})
		|0;1;0\rangle \\
		& = & -\int dt^{\prime}dt^{\prime\prime} d^3x^{\prime}
		d^3x^{\prime\prime}d\tilde{k} \epsilon_1(t^{\prime})
		\epsilon_2(t^{\prime\prime})
		h_1({\bf x}^{\prime}) 
		h_2^*({\bf x}^{\prime\prime}) \nonumber \\
		& & \;\;\times \exp\left[-i(E-\Omega_2)t^{\prime\prime}
			+i(E-\Omega_1)t^{\prime}
			+i{\bf k}\cdot ({\bf x}^{\prime\prime}-
			{\bf x}^{\prime}) \right].
\end{eqnarray}
Since the amplitude is proportional to 
$\langle 0|\phi(x^{\prime\prime})\phi^{\dagger}(x^{\prime})|0\rangle$,
it is clear from Eq.~(\ref{phimode}) that this interaction corresponds
to the creation and subsequent annihilation of a single neutrino.

In order to proceed further we choose $h_1$, $h_2$ and $\epsilon_1$
to be Gaussians since this allows many of the integrals to be
evaluated exactly.  Setting 
\begin{eqnarray}
	h_1({\bf x}) & = & \left( \sqrt{2\pi}\sigma_{x_1}\right)^{-3}
			e^{-|{\bf x}|^2/{2\sigma_{x_1}^2}} ,
	\label{psi1def}	\\
        h_2({\bf x}) & = & \left( \sqrt{2\pi}\sigma_{x_2}\right)^{-3}
                        e^{-|{\bf x}-{\bf x}_D|^2/{2\sigma_{x_2}^2}} ,
	\label{psi2def} \\
	\epsilon_1(t) & = & \epsilon_1^0 e^{-t^2/2\sigma_{t_1}^2}
	\label{eps1def}
\end{eqnarray}
we obtain
\begin{eqnarray}
	{\cal A} & = & -\left(\frac{\sqrt{2\pi}\epsilon_1^0\sigma_{t_1}}
			{4\pi^2 x_D}\right)
		\int_{-\infty}^{\infty} dt^{\prime\prime}
		\epsilon_2(t^{\prime\prime})
		\int_m^{\infty} dE 
		\exp\left[-i(E-\Omega_2)t^{\prime\prime}
                        -\frac{1}{2}(E - \Omega_1)^2\sigma_{t_1}^2
                \right. \nonumber \\
                & & \left. \;\; -\frac{1}{2}k^2(\sigma_{x_1}^2+
                                \sigma_{x_2}^2)\right]
		\sin(kx_D) ,
	\label{ampeps2}
\end{eqnarray}
 where 
\equation
	k\equiv \sqrt{E^2-m^2} .
\endequation

Before choosing an explicit form for $\epsilon_2(t)$,
which determines characteristics of the detector,
let us make a few observations regarding the 
above expression for the amplitude.
First of all, for large $x_D$, the amplitude 
decreases like $x_D^{-1}$
so that the probability falls like $x_D^{-2}$, as
expected on geometrical grounds in three dimensions.  
At the origin, however, the amplitude does not diverge (despite
the $1/x_D$ factor),
due to the sine function in the integrand.  A second
observation is that conservation of
energy at the source and of momentum at both the source and
detector are governed by the relative sizes of $\sigma_{t_1}$,
$\sigma_{x_1}$ and $\sigma_{x_2}$.  This situation is in accordance
with the uncertainty principle (and is in fact necessary,
as discussed above, in order to observe oscillations).
In general, neither energy nor momentum
need be conserved exactly if the source and detector are
localized in space and time.  The specific set-up which 
we have chosen favours energies close to the energy of the
excited source, $\Omega_1$, and momenta close to zero.
This latter point is due to the fact that our souce 
and detector have no dynamical degrees of freedom -- they cannot
recoil when a neutrino is emitted or absorbed -- and thus
the neutrino gets all of its momentum from the uncertainties
in the positions of the source and detector.  In order to
avoid the problem that low momenta are favoured,
we shall typically choose to 
set $\sigma_{t_1}$$\gg$$\sigma_{x_{1,2}}$ in
our numerical work below.\footnote{This is a ``trick'' which we
use to get sensible results, but it is also not unreasonable on
physical grounds.  According to the discussion in Ref.~\cite{knw},
for example, this condition is satisfied by several orders of magnitude
if $\sigma_x$ is taken to be on the order of nuclear sizes.  The
reader is also referred to the discussion of Lipkin~\cite{lipkin},
where this same point is emphasized.}
When several neutrino fields are coupled to the source and detector, this
will mean that the energies of the mass eigenstates will
be approximately equal, while their momenta will be determined
by their energies.  Furthermore, the sizes of the neutrino
wave packets will then be determined more by the amount of time
for which the source emits an uninterrupted wave-train than
by the localization of the source-field interaction in
configuration space.
In Sec.~\ref{sec3.3}, when we extend our analysis to fermionic
neutrinos, we will allow the source to decay by emitting
both a neutrino and its associated lepton.  In this case
the neutrino's momentum will no longer be centered about
$k$$\approx$$0$.

Let us now study the system as a function of the 
coherence time of the detector.  At one extreme we can 
imagine that a given (microscopic) detector is turned on for the
entire time that the neutrino ``wave packet'' passes by.  This
is an ultimately ``coherent'' detection event.  Another possibility
is that a given microscopic detector turns on and off without sampling
the entire wave packet.  In order to model the former scenario we
use a simple step function for $\epsilon_2$, while for the 
latter case we use a gaussian:
\begin{eqnarray}
	\epsilon_2^{\rm step}(t) & \equiv & \epsilon_2^0
		\theta(t_2-t)\theta(t-t_1), \label{eps2s}\\
	\epsilon_2^{\rm gauss}(t) & \equiv & \epsilon_2^0
		e^{-(t-t_D)^2/2\sigma_{t_2}^2}.
	\label{eps2g}
\end{eqnarray}
The step function detector turns on abruptly at time $t_1$ and
off again abruptly at time $t_2$ (with $t_1$ and $t_2$ chosen such that
the entire wave packet passes by while the detector is on), while the
gaussian detector turns on and off gradually at a time centered
around $t_D$.  Since the coherent (step function) detector
``catches'' the entire wave packet, there is no need to integrate
the resulting expression for the probability over time.  This is not
the case for the incoherent (gaussian) detector:  since each individual
microscopic detector sees only a piece of the wave packet, one must 
sum incoherently over all of the microscopic detectors in order to correctly
model the response of the bulk detector.  If there are many microscopic
detectors in the bulk detector, this incoherent sum corresponds to 
integrating the expression for the probability over $t_D$.

It is straightforward to evaluate the amplitudes for both 
types of detectors and we obtain
\begin{eqnarray}
        {\cal A}_{\rm step} & = & \tilde{N} (t_2-t_1)
                \int_m^{\infty} dE \,
                \frac{\sin\left[(E-\Omega_2)(t_2-t_1)/2\right]}
                {\left[(E-\Omega_2)(t_2-t_1)/2\right]} \,
                \exp\left[-\frac{i}{2}(E - \Omega_2)(t_1+t_2)
                \right. \nonumber \\
                & & \left. \;\; -\frac{1}{2}(E - \Omega_1)^2\sigma_{t_1}^2
                -\frac{1}{2}k^2(\sigma_{x_1}^2+
                                \sigma_{x_2}^2)\right]\sin(kx_D),
                \label{astep}\\
	{\cal A}_{\rm gauss} & = & \tilde{N} \sqrt{2\pi}\sigma_{t_2}
		\int_m^{\infty} dE
		\exp\left[-\frac{1}{2}(E - \Omega_1)^2\sigma_{t_1}^2
                        -\frac{1}{2}(E - \Omega_2)^2\sigma_{t_2}^2
                \right. \nonumber \\
                & & \left. \;\; -\frac{1}{2}k^2(\sigma_{x_1}^2+
                                \sigma_{x_2}^2)
			-i(E-\Omega_2)t_D \right]
		\sin(kx_D), 
		\label{agauss}
\end{eqnarray}
where 
\equation
	\tilde{N} \equiv -\frac{\epsilon_1^0\epsilon_2^0\sigma_{t_1}}
		{(2\pi)^{3/2}x_D} .
	\label{ndef}
\endequation
It is not possible in general to obtain analytic closed-form solutions
of these integrals, but they are simple to evaluate numerically.
In so doing, we obtain exact (to second order
in perturbation theory) solutions to the problem which we
are studying, including all effects due to the spreading of the 
neutrino wave packets.  Alternatively, we may, in some cases,
find reliable approximations for these integrals.  Such is
the case for the step function detector
if the neutrino's mass is not too close to the production
and detection thresholds ($m$$\ll$$\Omega_i - 1/\sigma_{t_1}$).
Taking $t_1$ to be a time before the first bit of neutrino
flux arrives at the detector and $t_2$ to be a time after
the entire neutrino wave packet has passed (formally, $t_2$$\to$$\infty$),
we find
\begin{eqnarray}
    \lim_{t_2\rightarrow\infty}{\cal A}_{\rm step}(x_D,t_1,t_2)
        & \simeq &
        -i\tilde{N}\pi\exp\left[i\bar{k}x_D
        -\frac{1}{2}(\Omega_2-\Omega_1)^2\sigma_{t_1}^2
        \right. \nonumber \\
        & & \left. \;\;\;\;
        -\frac{1}{2}(\Omega_2^2-m^2)
        (\sigma_{x_1}^2+\sigma_{x_2}^2)\right],
        \label{astepapprox}
\end{eqnarray}
where $\bar{k}$$\equiv$$(\Omega_2^2-m^2)^{1/2}$.  (The details of
this calculation may be found in Appendix~\ref{appendch3.2}.)
Note that the coherent detector ``picks out'' momenta
corresponding to the energy $\Omega_2$.                  

The modified ``probability'' associated with the above amplitude
is given by
\equation
        {\cal P}_{\rm step}(x_D) \equiv
                \left|{\cal A}_{\rm step}(x_D)\right|^2/
                \tilde{N}^2 ,
        \label{pstep}
\endequation
where we have dropped the implicit dependence on $t_1$ and $t_2$.
We have also divided through by $\tilde{N}^2$ because the value of that
constant (including the fall-off as $x_D^{-2}$) is not really
of interest to us since in any calculation of the oscillation probability
$\tilde{N}^2$ always factors out.  Taking the ratio of this
probability for two different values of the mass reveals that the 
system is more efficient at producing and detecting higher-mass
neutrinos:
\equation
        \frac{{\cal P}_{\rm step}(m;x_D)}
                {{\cal P}_{\rm step}(m^{\prime};x_D)}
                \simeq \exp\left[(m^2-m^{\prime 2})
                        (\sigma_{x_1}^2+\sigma_{x_2}^2)\right].
        \label{pstepfrac}
\endequation
The mass-dependence of the source/detector system arises due to
the fact that our source and detector favour neutrino
states with momenta close to zero.
This feature was predicted already in the discussion following
Eq.~(\ref{ampeps2}) and is due to the fact that the source and detector
in our model cannot ``recoil'' and thus the neutrino gets all of
its momentum from the uncertainty in the positions of the
source and detector.
Thus the upper limit on the neutrino's
momentum is given by $k_{\rm max}$$\sim$$1/\sigma_{x_{1,2}}$.
Note that this preference for non-relativistic
neutrinos is essentially a quirk of our model and
should not be viewed as a physical effect.
The mass-dependence of the system can be minimized
by setting $\sigma_{x_{1,2}}$ to be much less than
$\Omega_{1,2}^{-1}$.  In such
cases, the step function detector becomes nearly ``ideal;'' that is, it
detects neutrinos of different masses with nearly the same efficiency. 

We now turn to the gaussian detector and define
a modified probability in analogy with Eq.~(\ref{pstep})
\equation
        {\cal P}_{\rm gauss}(x_D,t_D,\sigma_{t_2}) \equiv
                \left|{\cal A}_{\rm gauss}(x_D,t_D,\sigma_{t_2})\right|^2/
		\tilde{N}^2.
        \label{pgauss}
\endequation
This expression gives the probability that a given microscopic
detector -- turned on for a time $\sigma_{t_2}$
centered around the time $t_D$ -- is excited.
We need to convert this expression into one giving the probability
that the bulk detector ``detects'' the neutrino (i.e., that one
of the micropscopic detectors is excited.)  We assume that the
bulk detector is ``on'' for all $t_D$$>$$0$ -- in the sense that
at any given time many of the microscopic detectors are 
``on'' -- but that the microscopic detectors themselves turn on and off
randomly, so that the number which are ``on'' at any given
time is roughly constant.  Then the probability that the bulk
detector ``detects'' the neutrino is proportional to the
integral of Eq.~(\ref{pgauss}) over $t_D$.\footnote{
Consider first a simpler case in which there are $N$ detectors,
turning on and off at times centered about
$t_1$$<$$t_2$$<$$\ldots$$<$$t_N$.  Each
of them has probability $\epsilon$ to detect the neutrino, but
{\em only} if one of the previous detectors has not already detected
it.  Then the probability that none of them detects the neutrino
is $(1-\epsilon)^N$, that the last one detects it is 
$(1-\epsilon)^{N-1}\epsilon$, that the second last one detects
it is $(1-\epsilon)^{N-2}\epsilon$, and so on.  The probabilities for
the $N+1$ distinct possibilities sum to unity, as required.
The probability that the neutrino is detected is then
$1-(1-\epsilon)^N$$=$$N\epsilon - N!\epsilon^2/(N-2)!2!
+\ldots$$\simeq$$N\epsilon$ if $N\epsilon$$\ll$$1$, that is,
if the probability of detecting the neutrino in the bulk
detector is much less than one (which is
certainly the case).  In the case at hand suppose 
that $t_1$ corresponds to a time before any appreciable flux has
arrived at the detector and $t_N$$=$$t_1+T$ to a time after all of the flux
has passed.  Then 
\begin{eqnarray}
	\sum_{i=1}^N {\cal P}(x_D,t_i) & \equiv &
	\frac{(N-1)}{T}\sum_{i=1}^N {\cal P}(x_D,t_i)\Delta t\simeq 
	\frac{(N-1)}{T}\int_{t_1}^{t_1+T} dt_D
	{\cal P}(x_D,t_D)\nonumber .
\end{eqnarray}
}  
We thus refer to this type of bulk detector as an ``incoherent''
detector, since we sum the probability incoherently over different times.

The time integral of Eq.~(\ref{pgauss}) may actually be done
explicitly.  Let us define the following
(unnormalized) time-integrated probability 
\equation
    {\cal P}_{\rm incoh}(x_D,\sigma_{t_2}) \equiv 
            \int_0^{\infty} dt_D{\cal P}_{\rm gauss}(x_D,t_D,\sigma_{t_2}) .
	\label{ptimeint}
\endequation
Since the integrand is symmetric under $t_D$$\rightarrow$$-t_D$,
we may formally extend the integration to negative infinity and divide by two.
The time integral then reduces to a delta function in energy and allows
us to perform one of the energy integrals.  As a result, we
obtain
\begin{eqnarray}
    {\cal P}_{\rm incoh}(x_D,\sigma_{t_2}) & = &
            2\pi^2\sigma_{t_2}^2\int_m^{\infty} dE 
		\exp\left[-(E - \Omega_1)^2\sigma_{t_1}^2
                        -(E - \Omega_2)^2\sigma_{t_2}^2
                \right. \nonumber \\
                & & \left. \;\; -k^2(\sigma_{x_1}^2+
                                \sigma_{x_2}^2)\right]
		\sin^2(kx_D).
	\label{pincoh}
\end{eqnarray}

If $m$$\ll$$\Omega_i-1/\sigma_{t_{i}}$ and 
$\sigma_{t_{1,2}}$$\gg$$\sigma_{x_{1,2}}$
(the latter condition is always assumed) 
then we may approximate the above expression
by setting $\sin^2(kx_D)$$\approx$$1/2$ to yield
\begin{eqnarray}
	{\cal P}_{\rm incoh}(x_D,\sigma_{t_2}) & \simeq &
		\frac{\pi^{5/2}\sigma_{t_2}^2}
		{\left(\sigma_{t_1}^2+\sigma_{t_2}^2
		+\sigma_{x_1}^2+\sigma_{x_2}^2\right)^{1/2}}
		\exp\left[m^2(\sigma_{x_1}^2+\sigma_{x_2}^2)\right]
		\nonumber \\
		& & \;\;\times \exp\left[\frac{
		\left(\Omega_1\sigma_{t_1}^2+\Omega_2\sigma_{t_2}^2\right)^2}
		{\left(\sigma_{t_1}^2+\sigma_{t_2}^2
                +\sigma_{x_1}^2+\sigma_{x_2}^2\right)}
		-\Omega_1^2\sigma_{t_1}^2-\Omega_2^2\sigma_{t_2}^2\right].
	\label{pincohapprox}
\end{eqnarray}
Thus, under the above conditions the ``incoherent'' 
gaussian detector has the same mass-dependence as the
step function detector does (c.f. Eq.~(\ref{pstepfrac})):
\equation
        \frac{{\cal P}_{\rm incoh}(m;x_D,\sigma_{t_2})}
                {{\cal P}_{\rm incoh}(m^{\prime};x_D,\sigma_{t_2})}
                \simeq \exp\left[(m^2-m^{\prime 2})
                        (\sigma_{x_1}^2+\sigma_{x_2}^2)\right].
        \label{pgaussfrac}
\endequation
This fact is rather remarkable and shows again that it is correct
to perform the time integral in Eq.~(\ref{ptimeint}).

\begin{figure}[tb]
\epsfbox[68 395 502 714]{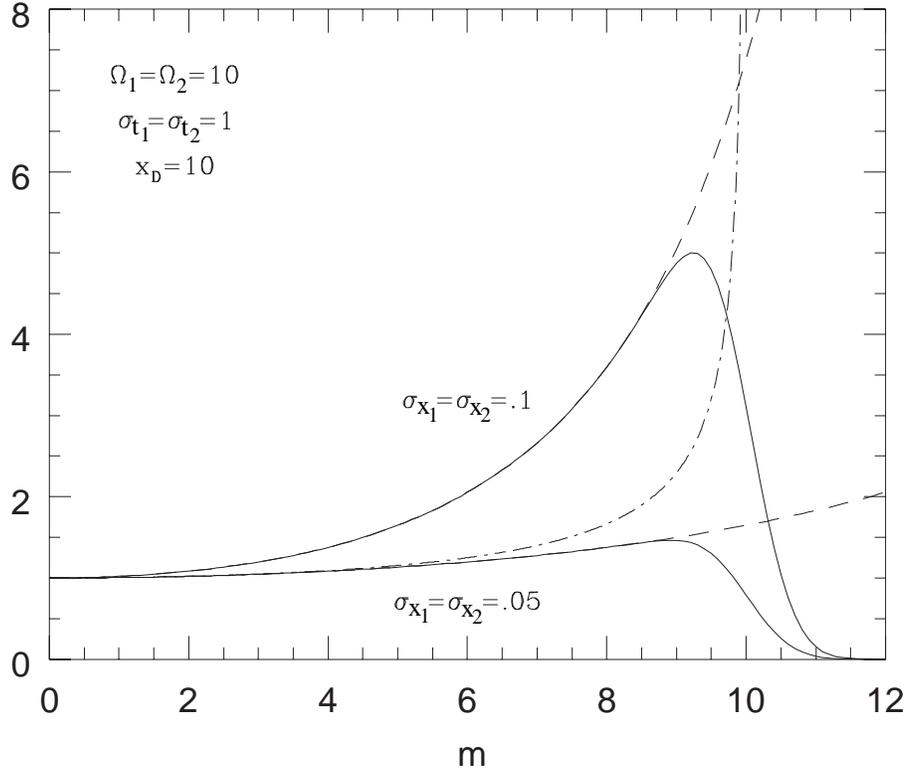}
\caption[Mass dependence of the gaussian detector]
{Plot of the ``incoherent probability'' ${\cal P}_{\rm incoh}$
(Eq.~(\ref{pincoh}), normalized to its value at $m$$=$$0$)
as a function of mass
for the case of a single neutrino,
taking $\sigma_{x_{1,2}}$$=$$0.1, 0.05$.  In each case,
the solid line shows the exact result and the dashed line
shows the result obtained in the approximation of
Eq.~(\ref{pincohapprox}).  The dash-dotted line shows a plot
of $1/v(m)$ for comparison.}
 \label{prob3.fg}
\end{figure}

Fig.~\ref{prob3.fg} shows a plot of the time-integrated
probability, Eq.~(\ref{pincoh}), as a function of the neutrino
mass for two different values of $\sigma_{x_{1,2}}$.
This probability may be regarded as giving a measure
of the {\em efficiency} with which the system produces and detects
a neutrino of a given mass.
For convenience, the probabilities have been normalized
to their values at $m$$=$$0$.
In each case,
the solid line gives the exact result and the dashed line
shows the approximation for non-threshold
masses derived in Eq.~(\ref{pincohapprox}).
Clearly the approximation is quite good if the mass is
not too close to the neutrino production and detection
thresholds.  Furthermore, it is clear that this detector
can be made ``ideal'' (that is, the probability to
detect a neutrino may be
made mass-independent) by using suitably small values for 
$\sigma_{x_{1,2}}$.  The mass-dependence for large
$\sigma_{x_{1,2}}$ occurs for the same
reason as in the case of the step function detector and 
is due to the fact that
the source and detector in our model cannot recoil (see
the discussion following Eq.~(\ref{pstepfrac})).
The dash-dotted line shows a plot of $1/v(m)$ for comparison.
This would be the analogous efficiency found in a wave packet 
calculation~\cite{kimpacket1}.  Since our detector may be
made mass-independent by a suitable choice of the $\sigma_{x_i}$,
we see that a ``well designed'' detector will not exhibit
such an effect.

\subsection{Several Neutrinos}
\label{sevneut}

Now that we have studied the characteristics of the source/detector
system in the single-neutrino case, we turn to the case in 
which there are several neutrino fields coupled to the source
and detector.  Suppose that there are $N$ different
neutrino mass eigenstates.  Then, in order to model the real-life
situation, we suppose that there are also several different types 
of sources and detectors, each of which couple to a
``weak eigenstate'' which is a given 
unitary linear combination of the neutrino mass eigenstates.
The action of Eq.~(\ref{toyaction}) is then generalized to
\equation
        S = \int d^4x \left({\cal L}_{\phi}^0 + {\cal L}_{\rm int}\right)
                + \int dt L_q^0,
\endequation
where
\begin{eqnarray}
        {\cal L}_{\phi}^0 & = & 
                -\sum_i\phi^{\dagger}_i(x)
		\left(\Box + m^2_i\right)\phi_i(x), \\
        L_q^0 & = & \sum_{\alpha}\left[\dot{q}_1^{\alpha\dagger}(t)
		\dot{q}_1^{\alpha}(t) 
                - \Omega_1^{\alpha 2}q_1^{\alpha\dagger}(t)q_1^{\alpha}(t)
                +\dot{q}_2^{\alpha\dagger}(t)\dot{q}_2^{\alpha}(t) 
                - \Omega_2^{\alpha 2}q_2^{\alpha\dagger}(t)q_2^{\alpha}(t)
		\right] ,\\
        {\cal L}_{\rm int} & = & -\sum_{\alpha, i}\left[
                \epsilon_1(t)\left({\cal U}_{\alpha i}^*\phi^{\dagger}_i(x)
			q_1^{\alpha}(t)h_1({\bf x})+
                        {\cal U}_{\alpha i}\phi_i(x)q_1^{\alpha\dagger}(t)
			h_1^*({\bf x})\right)\right.
                \nonumber \\
                & & \left. \;\;\;\;\;\; +
                \epsilon_2(t)\left({\cal U}_{\alpha i}^*\phi^{\dagger}_i(x)
			q_2^{\alpha}(t)h_2({\bf x})+
                        {\cal U}_{\alpha i}\phi_i(x)q_2^{\alpha\dagger}(t)
			h_2^*({\bf x})\right)\right] ,
\end{eqnarray}
and in which ${\cal U}$ is a unitary matrix.  Note that the
subscripts ``1'' and ``2'' on the functions $\epsilon$ and $h$
and on the fields $q$
refer, respectively, to the source and detector.  These should not
be confused with the subscripts on the fields $\phi$ which refer
to the mass eigenstates.  Also note that we have taken
$\epsilon$ and $h$ to be independent of the flavour or
mass eigenstate in question.  In principle there could be 
such a dependence, but including it would unnecessarily
complicate our analysis.  In what follows, we shall also
set $\Omega_i^{\alpha}$$=$$\Omega_i$, for all $\alpha$, in 
order to ``idealize'' our sources and detectors.

The experimental set-up which we wish
to consider is a simple generalization of that given
in the previous section.  In this case we imagine that
the initial state of the system has an $\alpha$-flavour ``source''
oscillator in its first excited state and that the final state has
a $\beta$-flavour ``detector'' oscillator in its first excited state.
The amplitude for this process may then be calculated as in
the single-neutrino case and we find
\begin{eqnarray}
        {\cal A}_{\alpha\rightarrow\beta} & = & 
		-\sum_{i,j}{\cal U}_{\beta j}{\cal U}_{\alpha i}^*
		\langle 0;0;1_{\beta}|\int dt^{\prime}
                dt^{\prime\prime} d^3x^{\prime}
                d^3x^{\prime\prime} \epsilon_1(t^{\prime})
                \epsilon_2(t^{\prime\prime}) \nonumber \\
                & & \;\;\times
                \phi_j(x^{\prime\prime})q_2^{\beta\dagger}(t^{\prime\prime})
                h_2^*({\bf x}^{\prime\prime})
                \phi^{\dagger}_i(x^{\prime})
                q_1^{\alpha}(t^{\prime})h_1({\bf x}^{\prime})
                |0;1_{\alpha};0\rangle \\
                & = & -\sum_i {\cal U}_{\beta i}{\cal U}_{\alpha i}^*
		\int dt^{\prime}dt^{\prime\prime} d^3x^{\prime}
                d^3x^{\prime\prime}d\tilde{k}_i \epsilon_1(t^{\prime})
                \epsilon_2(t^{\prime\prime})
                h_1({\bf x}^{\prime}) 
                h_2^*({\bf x}^{\prime\prime}) \nonumber \\
                & & \;\;\times \exp\left[
			-i(E_i-\Omega_2)t^{\prime\prime}
                        +i(E_i-\Omega_1)t^{\prime}
                        +i{\bf k}\cdot ({\bf x}^{\prime\prime}-
                        {\bf x}^{\prime}) \right] ,
\end{eqnarray}
in which we have defined
\begin{eqnarray}
	E_i & \equiv & \sqrt{k^2 + m_i^2} .
\end{eqnarray}
Taking $h_1$, $h_2$ and $\epsilon_1$ to be
gaussians with widths $\sigma_{x_1}$, $\sigma_{x_2}$ and $\sigma_{t_1}$
as in the single-neutrino case
(see Eqs.~(\ref{psi1def}), (\ref{psi2def}) and 
(\ref{eps1def})), we may further simplify this expression
\begin{eqnarray}
        {\cal A}_{\alpha\rightarrow\beta} 
		& = & -\left(\frac{\sqrt{2\pi}\epsilon_1^0\sigma_{t_1}}
                        {4\pi^2 x_D}\right)
		\sum_i{\cal U}_{\beta i}{\cal U}_{\alpha i}^*
                \int_{-\infty}^{\infty} dt^{\prime\prime}
                \epsilon_2(t^{\prime\prime})
               \int_{m_i}^{\infty} dE 
                \exp\left[-i(E-\Omega_2)t^{\prime\prime}
                \right. \nonumber \\
		& & \left. \;\;
		-\frac{1}{2}(E - \Omega_1)^2\sigma_{t_1}^2
                -\frac{1}{2}k_i^2(\sigma_{x_1}^2+
                                \sigma_{x_2}^2)\right]
                \sin(k_ix_D) ,
\end{eqnarray}
where 
\equation
        k_i\equiv \sqrt{E^2-m_i^2} .
\endequation
This expression is clearly just a simple generalization of
Eq.~(\ref{ampeps2}).  The final step in our calculation is
to substitute the expressions (\ref{eps2s}) and (\ref{eps2g})
for $\epsilon_2(t)$ in the step function and gaussian
detector cases.  This yields
\begin{eqnarray}
        {\cal A}^{\rm step}_{\alpha\rightarrow\beta}
		 & = & \tilde{N} (t_2-t_1)
		\sum_i{\cal U}_{\beta i}{\cal U}_{\alpha i}^*
                \int_{m_i}^{\infty} dE \,
                \frac{\sin\left[(E-\Omega_2)(t_2-t_1)/2\right]}
                {\left[(E-\Omega_2)(t_2-t_1)/2\right]} \,
		\sin(k_ix_D)  \nonumber \\
		& &  \times\exp\left[
		-\frac{i}{2}(E - \Omega_2)(t_1+t_2)
                -\frac{1}{2}(E - \Omega_1)^2
		\sigma_{t_1}^2 -\frac{1}{2}k_i^2(\sigma_{x_1}^2+
                                \sigma_{x_2}^2)\right],
                \label{astep2}\\
        {\cal A}^{\rm gauss}_{\alpha\rightarrow\beta}
                 & = & \tilde{N} \sqrt{2\pi}\sigma_{t_2}
                \sum_i{\cal U}_{\beta i}{\cal U}_{\alpha i}^*
                \int_{m_i}^{\infty} dE
                \exp\left[-\frac{1}{2}(E - \Omega_1)^2\sigma_{t_1}^2
                        -\frac{1}{2}(E - \Omega_2)^2\sigma_{t_2}^2
                \right. \nonumber \\
                & & \left. \;\; -\frac{1}{2}k_i^2(\sigma_{x_1}^2+
                                \sigma_{x_2}^2)
                        -i(E-\Omega_2)t_D \right]
                \sin(k_ix_D),
                \label{agauss2}
\end{eqnarray}
where $\tilde{N}$ is defined in (\ref{ndef}).  

We are finally in a position to define the oscillation probability
as a function of distance for the two cases.  In both
cases our definition of the probability is a ``physical'' one.
We imagine that the source produces neutrinos of type $\alpha$
($\alpha$$=$$e,\mu,\tau,\ldots$) and that
we set a $\beta$-neutrino detector at some distance $x_D$ from the 
source.  We prepare the source (or an ensemble of identically
prepared sources) in an excited state, wait a long period
of time, and then check to see if the detector has been excited.
After repeating this experiment enough times to get good statistics,
we repeat the procedure with a $\beta^{\prime}$-neutrino detector, and so
on.  The probability to observe a $\beta$ neutrino is then simply
the number of events observed in ``$\beta$-mode'' divided by the 
total number of events in all modes.  Since we have attempted to 
make our source/detector system as ``ideal'' as possible, there
are no further corrections for detector efficiencies or other effects
of that nature.  The normalized coherent and incoherent 
oscillation probabilities may then be defined as
\begin{eqnarray}
	{\cal P}_{\alpha\rightarrow\beta}^{\rm coh} (x_D)
	& = & \lim_{t_2\rightarrow\infty} 
		\frac{\left|{\cal A}^{\rm step}_{\alpha\rightarrow\beta}
			(x_D,t_1,t_2)\right|^2}
		{\sum_{\beta^\prime}\left|
		{\cal A}^{\rm step}_{\alpha\rightarrow\beta^\prime}
                        (x_D,t_1,t_2)\right|^2} , 
	\label{pcohab}	\\
        {\cal P}_{\alpha\rightarrow\beta}^{\rm incoh} (x_D,\sigma_{t_2})
	& = & \frac{ \int_0^{\infty} dt_D 
		\left|{\cal A}^{\rm gauss}_{\alpha\rightarrow\beta}
                        (x_D,t_D,\sigma_{t_2})\right|^2}
                {\sum_{\beta^\prime} \int_0^{\infty} dt_D
                \left|{\cal A}^{\rm gauss}_{\alpha\rightarrow\beta^\prime}
                        (x_D,t_D,\sigma_{t_2})\right|^2} .
	\label{pincohab}
\end{eqnarray}	
It is understood in the first expression that $t_1$ 
is taken to be some time before the first bit of neutrino
``flux'' arrives at the detector.

The expressions which we have derived for our two types of
detectors are in forms which are amenable to numerical calculation.
The coherent probability may be found after a single integration
over energy and the incoherent probability requires two integrations,
one over energy and one over time.  In the two-neutrino case, 
the time integral in Eq.~(\ref{pincohab}) may be done by
hand, but this is not possible in general for more neutrinos.
The reason for this is that the integrand is no longer
symmetric under $t_D$$\rightarrow$$-t_D$ due to the possible presence of
phases in the mixing matrix ${\cal U}$.

Let us examine the case with two flavours in some detail.  In this case,
the matrix ${\cal U}$ may be taken to be a real orthogonal matrix
parametrized by one angle, $\theta$.  The time integral in the numerator
of (\ref{pincohab}) may be performed explicitly and
we find
\begin{eqnarray}
	\int_0^{\infty} dt_D
                \left|{\cal A}^{\rm gauss}_{\alpha\rightarrow\beta}
                (x_D,t_D,\sigma_{t_2})\right|^2 & = &  
		\nonumber \\
		& & \hspace{-2.25in} 
		2\pi^2\sigma_{t_2}^2\tilde{N}^2 
		\sum_{i,j}{\cal U}_{\beta i}{\cal U}_{\beta j}
		{\cal U}_{\alpha i}{\cal U}_{\alpha j}
		\int_{{\rm max}(m_i,m_j)}^{\infty} dE 
		\sin(k_ix_D)\sin(k_jx_D) \nonumber \\
		& & \hspace{-2.25in} \times
                \exp\left[-(E - \Omega_1)^2\sigma_{t_1}^2
                        -(E - \Omega_2)^2\sigma_{t_2}^2
                 -(E^2-(m_i^2+m_j^2)/2)(\sigma_{x_1}^2+
                                \sigma_{x_2}^2)\right] .
	\label{sigtsym}
\end{eqnarray}

\begin{figure}[htbp]
\epsfbox[60 193 481 696]{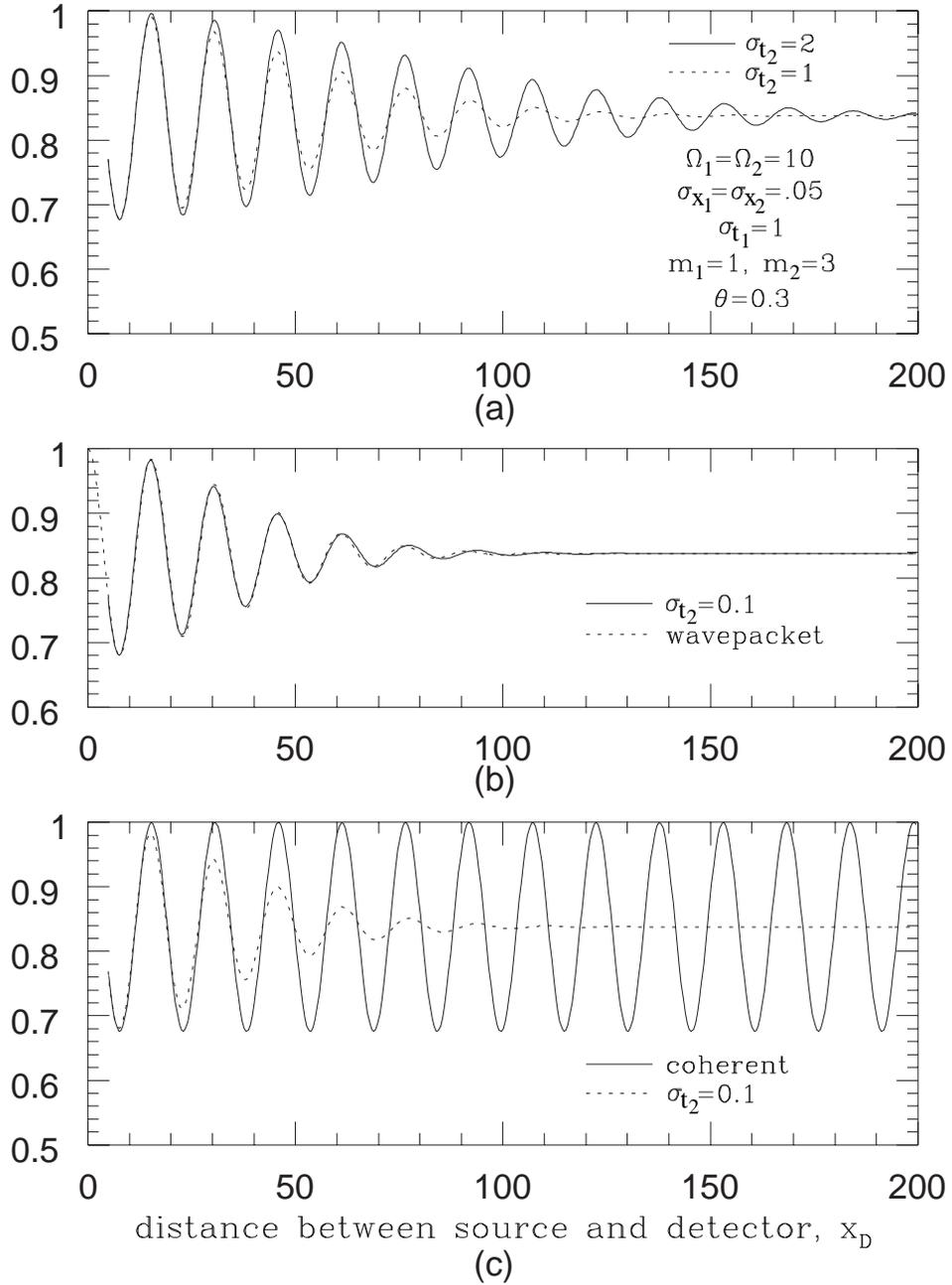}
\caption[Oscillation probabilities as a function of distance]
{Oscillation probabilities as a function of distance.  The 
two curves in (a) correspond to the ``incoherent'' detector
with time resolutions $\sigma_{t_2}$$=$$1, 2$.  The solid
curve in (b) gives the ``incoherent'' probability for 
$\sigma_{t_2}$$=$$0.1$.  The dotted curve shows the
analogous result obtained in the wave packet approach.
The solid curve in (c) shows the probability measured
by the ``coherent'' detector.}
\label{prob4.fg}
\end{figure}

Fig.~\ref{prob4.fg} shows several plots of the flavour-conserving
probability ${\cal P}_{e\rightarrow e}(x_D)$ as a function of $x_D$
for two relativistic neutrinos, using both the ``coherent'' and
the ``incoherent'' detector.  The various parameters chosen
for the plot are as indicated in the figure.  
Recall that $\Omega_1$ and $\Omega_2$ (set equal here) are the
energies of the excited source and detector, respectively.
Since we have chosen to set $\sigma_{x_i}$$\ll$$\sigma_{t_i}$, 
the energies of the mass eigenstates are approximately equal
to $\Omega$ and their momenta are determined by their 
energies.  The values
employed here for $\theta$, $m_1$ and $m_2$ are chosen
merely for the purpose of illustration.  Note that
the curves do not go all the way to $x_D$$=$$0$, since
our formulas are not valid for very small $x_D$.\footnote{Recall 
that we require the source to turn off
before the detector turns on in order that we may drop one
of the time-orderings in the neutrino propagator.
The reader is referred to the discussion following Eq.~(\ref{adefined})
for more details on this point.}

Figs.~\ref{prob4.fg}(a) and (b) show plots of the 
probability for detecting the same-flavour neutrino as was emitted
in the case of an ``incoherent'' detector (see Eq.~(\ref{pincohab})) 
for several different values
of the time resolution of the detector, $\sigma_{t_2}$.  The dotted
curve in (b) is the analogous result derived using the wave packet
approach~\cite{kimpacket1}.  This appears to be a good
approximation to our result in the limit as 
$\sigma_{t_2}$$\rightarrow$$0$.  It is
clear from these plots that the coherence length of the
oscillations is dependent on the time resolution of the
detector; that is, as was noted in Ref.~\cite{knw},
a long coherent measurement in time is capable of
``reviving'' oscillations of neutrinos whose mass eigenstate
wave packets have become physically separated.  This
effect is particularly striking in the case of the probability
detected by the coherent detector, shown 
by the solid curve in Fig.~\ref{prob4.fg}(c).
In this case the oscillations appear to have been {\em completely
revived} even after, according to an ``incoherent'' measurement
(dotted curve), the wave packets have completely separated.

\begin{figure}[htbp]
\epsfbox[60 236 489 708]{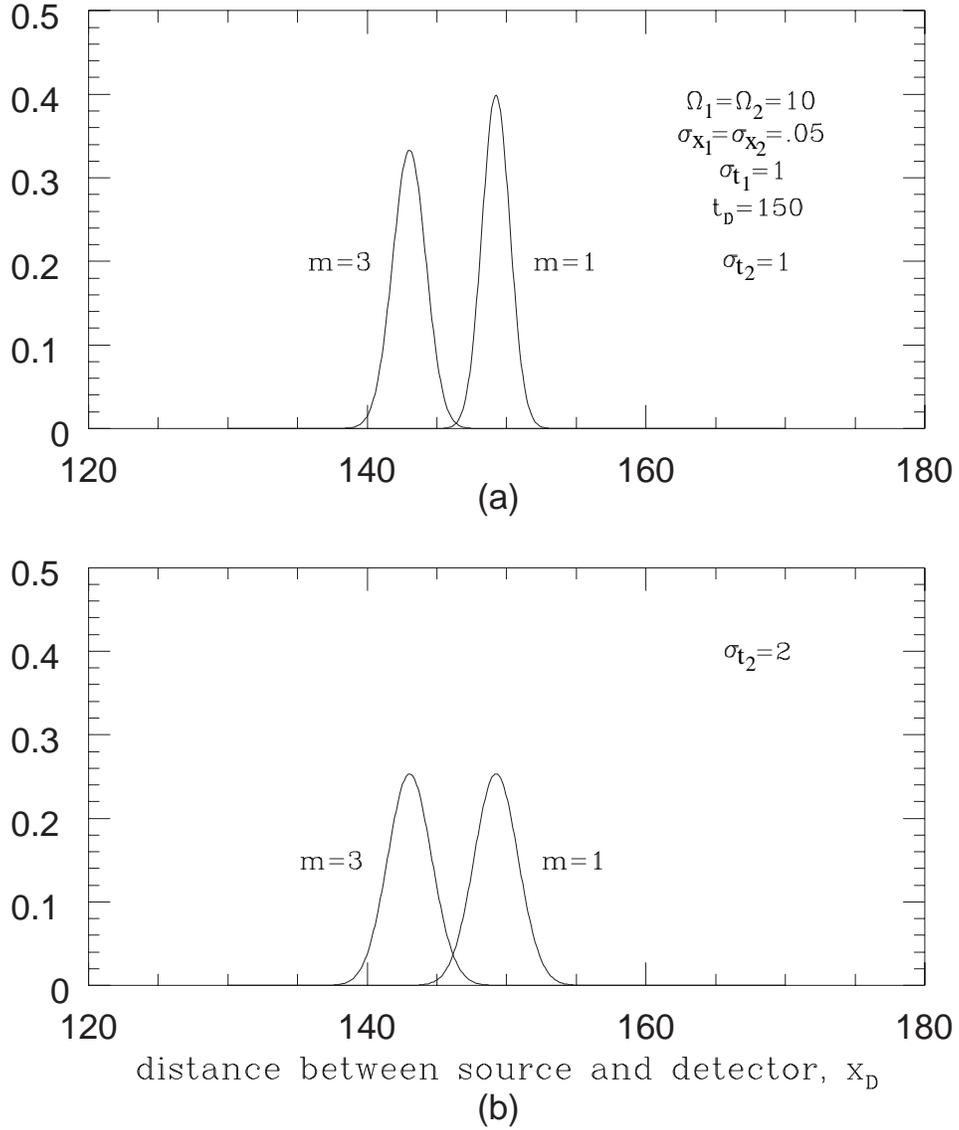}
\caption[Separation of wave packets for different time resolutions]
{``Snapshots'' of two mass eigenstate wave packets using 
incoherent detectors with different time resolutions.  The wave packets
have been individually normalized over $x_D$.  In (a) the time
resolution of the detector is such that the wave packets appear
to be nearly separated, while in (b) the same wave packets appear
to overlap due to the broader temporal resolution in that case.}
\label{prob5.fg}
\end{figure}

We have already discussed to some extent
how it is possible for a long coherent measurement in time
to revive the oscillations of neutrinos even after
the mass eigenstates have separated spatially.
Essentially, the accurate measurement of the energy
picks out the plane wave in the wave packet which has 
existed coherently through both pulses~\cite{knw}.
Our present approach allows for a complementary way
to view the situation.    
The question of whether the wave packets corresponding
to two mass eigenstates have separated or not
depends on the temporal and spatial resolution
of the detector.
We may demonstrate this effect by way of an example.
Let us take a ``snapshot'' of the wave packets 
corresponding to two different mass eigenstates at a fixed
time $t_D$$=$$150$ using the incoherent detector with different
widths, $\sigma_{t_2}$.  Figs.~\ref{prob5.fg}(a) and (b) show 
the detection probabilities (given by Eq.~(\ref{pgauss}), 
but separately normalized over $x_D$) for the two mass
eigenstates.  In (a), the time resolution of the detector
is taken to be $\sigma_{t_2}$$=$$1$ and there is 
almost no overlap between the two wave packets.  Indeed, comparison
with Fig.~\ref{prob4.fg}(a) shows that, for $x_D$$\approx$$150$,
the oscillations have been almost completely damped out.
If the detector is taken to have a broader time
resolution as in Fig.~\ref{prob5.fg}(b), however, the wave packets appear
to have a non-negligible overlap.  In this case
the width due to the finite time resolution of the detector has been 
added to the original widths of the wave packets.
Comparison with Fig.~\ref{prob4.fg}(a) shows that in
this case the oscillations have not yet been wiped out
for $x_D$$\approx$$150$.  From this point of view, then, the
fact that the conventional ``wave packet'' approach for
relativistic neutrinos agrees
with the source/detector approach (see Fig.~\ref{prob4.fg}(b))
for very small $\sigma_{t_2}$ is not that surprising.  The
wave packet approach simply ignores the finite time
resolution of the detector.

The ``coherent'' and ``incoherent'' probabilities, Eqs.~(\ref{pcohab})
and (\ref{pincohab}), may both be reliably approximated in the relativistic
limit.  Setting $\Omega$$\equiv$$\Omega_1$$=$$\Omega_2$ for
convenience, 
we obtain\footnote{We have used the approximate form of
${\cal A}^{\rm step}$ given in Eq.~(\ref{astepapprox})
in order to derive Eq.~(\ref{pcohabapprox}).  Also, in
deriving Eq.~(\ref{eq251}) we have 
dropped the highly oscillatory terms in the integrand
since they are strongly damped for $x_D$$>$$\sigma_{t_1}+\sigma_{t_2}$.
Recall that our calculation is only sensible for 
$x_D$$>$$\sigma_{t_1}+\sigma_{t_2}$.}
\equation
        {\cal P}_{\alpha\rightarrow\beta}^{\rm coh} (x_D) \simeq
		\frac{1}{\cal N}
                \sum_{i,j}{\cal U}_{\beta i}{\cal U}_{\alpha i}
                   {\cal U}_{\beta j}{\cal U}_{\alpha j}
                   \exp\left[i(\bar{k}_i-\bar{k}_j)x_D
                        +(m_i^2+m_j^2)(\sigma_{x_1}^2+\sigma_{x_2}^2)/2
                        \right] 
        \label{pcohabapprox}
\endequation
and
\begin{eqnarray}
        {\cal P}_{\alpha\rightarrow\beta}^{\rm incoh}
                (x_D,\sigma_{t_2}) & \simeq &
                \frac{1}{\cal N}
                \sum_{i,j}{\cal U}_{\beta i}{\cal U}_{\alpha i}
                   {\cal U}_{\beta j}{\cal U}_{\alpha j}
                   \exp\left[i(\bar{k}_i-\bar{k}_j)x_D
                        +(m_i^2+m_j^2)(\sigma_{x_1}^2+\sigma_{x_2}^2)/2
                        \right] \nonumber \\
        & & \;\;\;\;\;\;\;\;\;\;\; 
		\times \exp\left[-\frac{x_D^2\left(1/v_i
                        -1/v_j\right)^2}{4\left(\sigma_{t_1}^2
                        +\sigma_{t_2}^2\right)}\right]
        \label{eq251}
\end{eqnarray}
for the coherent and incoherent cases, respectively, where
we have defined
\begin{eqnarray}
	{\cal N} & = & \sum_{i}{\cal U}_{\alpha i}^2
                   \exp\left[m_i^2(\sigma_{x_1}^2+\sigma_{x_2}^2) ,
                        \right] \\
	\bar{k}_i & = & \sqrt{\Omega^2-m_i^2} , \\
	v_i & = & \bar{k}_i/\Omega .
\end{eqnarray}
These expressions are identical except for the
damping of the cross-terms which occurs
in the approximation for the ``incoherent'' case, Eq.~(\ref{eq251}).
Note that the oscillation length which may be extracted
 from either of these expressions is exactly what
one finds in the usual approach, with no spurious
factor of ``$2$.''
The approximation for the ``coherent'' case contains
no damping whatsoever, demonstrating that an
infinitely long coherent measurement does indeed completely
revive the oscillations of the neutrinos!
We also note that, while our expression for the 
incoherent case, Eq.~(\ref{eq251}), bears some
resemblance to the analogous expression obtained in the wave packet
approach (see, for example,~\cite{kimpacket1}), our expression
has an intrinsic dependence on the temporal and spatial resolution
of the detector which is ignored in the wave packet approach.
(This dependence on the detection process has also been investigated recently
in ~\cite{gkl97}.)
Finally, we note the absence of factors of $1/v_i$
pre-multiplying the exponentials such as can occur in
wave packet calculations.

\subsection{The Non-relativistic Case}
\label{sec:nonrelcase}

It is worthwhile to consider briefly the oscillations of non-relativistic
neutrinos in our toy model.  Let us assume that one of the mass eigenstates
is relatively light and let us study the behaviour of the oscillation
probability as the mass of the other neutrino (in the two-neutrino case)
is varied.  Furthermore, let us restrict our attention to the 
case of the incoherent detector, which is the more realistic of the
two detector types.  As the mass of the heavier neutrino increases, the packets
separate more quickly and, for sufficiently non-relativistic neutrinos,
the oscillations are damped out almost immediately.  It is convenient,
then, to simply study the asymptotic expression
\equation
	{\cal P}_{\alpha\rightarrow\beta}^{\infty} (\sigma_{t_2})
		\equiv	\lim_{x_D\rightarrow\infty} 
		{\cal P}_{\alpha\rightarrow\beta}^{\rm incoh} 
			(x_D,\sigma_{t_2}) .
	\label{limxinfpincoh}
\endequation
The main non-relativistic effect in our toy model is  due
to the model's dependence on $\sigma_{x_{1,2}}$
rather than being related to nonrelativistic effects of the
oscillations themselves. Recall from our
discussion in Sec.~\ref{asingleneut} that our source and detector
are more efficient at producing and detecting non-relativistic
neutrinos (see also Fig.~\ref{prob3.fg}.)  This dependence
skews the results for the oscillations, as one might expect.

\begin{figure}[tb]
\epsfbox[60 394 489 708]{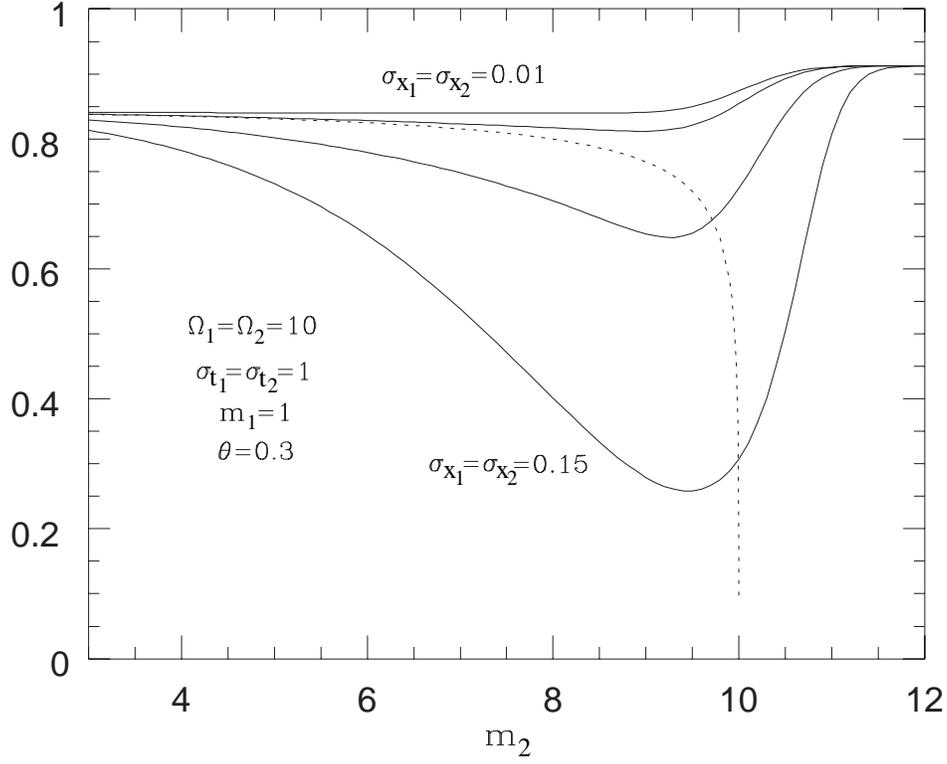}
\caption[Constant flavour-conserving probability as a function of $m_2$]
{Plot of the constant flavour-conserving probability,
${\cal P}^{\infty}_{e\rightarrow e}$, defined in
Eq.~(\ref{limxinfpincoh}) as a function of the mass of the heavier
neutrino.  The solid lines correspond to spatial widths
$\sigma_{x_{1,2}}$$=$$0.01, 0.05, 0.1, 0.15$ and the dotted line
shows the value obtained in the wave packet approach if the 
contributions corresponding to the various mass eigenstates are
weighted by $1/v_i$.}
\label{prob6.fg}

\end{figure}

In Fig.~\ref{prob6.fg} we have plotted the 
probability for a $\nu_e$ to be detected as a $\nu_e$
in the limit as $x_D$$\rightarrow$$\infty$
(${\cal P}^{\infty}_{e\rightarrow e}$
in Eq.~(\ref{limxinfpincoh}))
as a function of the mass of the heavier
neutrino.  The various curves
correspond to different values of $\sigma_{x_{1,2}}$,
the spatial resolution of the source and detector.  For
larger values of $\sigma_{x_{1,2}}$, this probability is indeed
skewed quite dramatically due to the fact
that the heavier mass eigenstate starts to dominate the
probability distribution.  Recall our earlier explanation
as to why this occurs in our model.  Since our
source and detector do not ``recoil'' when the neutrino
is emitted or absorbed, the upper limit on the 
neutrino's momentum is given by
$k_{\rm max}$$\sim$$1/\sigma_{x_{1,2}}$ (the reader is referred to the
discussion following Eq.~(\ref{pstepfrac})).
We emphasize, however, that this
effect is an artifact of our model and would not be expected to
occur in more realistic models.  We shall discuss
this point further below when we consider how our approach 
might be extended to the more realistic case in which the neutrino
is not the only decay particle emitted.  
Note also that as $m_2$ increases above the production/detection
threshold all of the solid curves approach the same value
of $\cos^2\theta$.  (How abrupt the threshold is depends on
how large $\sigma_{t_1}$ and $\sigma_{t_2}$ are, of course.)
The dotted curve shows, for comparison, the result which is obtained
if the detection efficiencies for the mass eigenstates are weighted
by $1/v_i$.  We see no evidence
in our model for this type of behaviour.

\section{Towards a More Realistic Calculation}
\label{sec3.3}

\begin{figure}[tb]
\epsfbox[95 374 529 712]{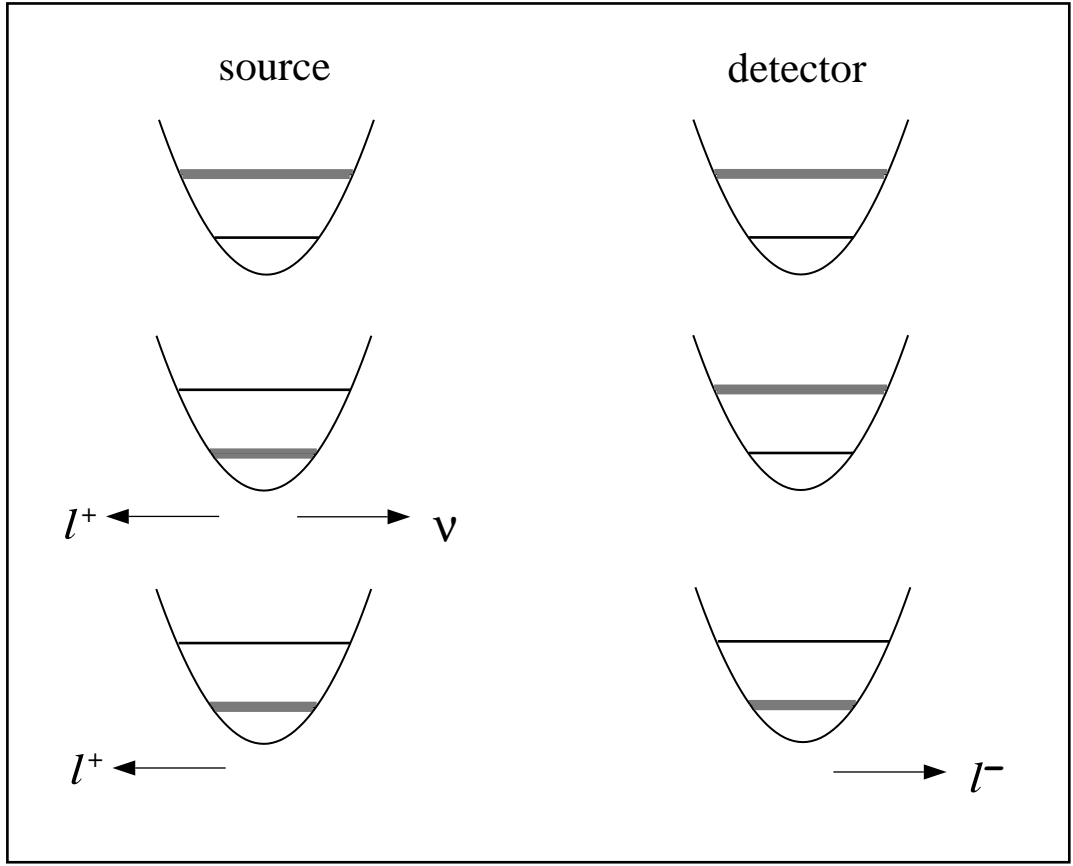}
\caption[Schematic illustration of the source/detector system]
{A schematic illustration of the sequence of events in the
source/detector system for fermionic neutrinos
considered in Sec.~\ref{sec3.3}.  The excited source
decays by emitting a neutrino and its associated anti-lepton.
The detector subsequently absorbs the neutrino and emits a lepton.}
\label{spin.fg}
\end{figure}

In this section we show how the bosonic model of the previous
section may be modified to account correctly for the 
fermionic nature of the neutrinos (which we shall assume to be
Dirac neutrinos) and for the $V-A$ nature
of neutrino interactions. This will have the added benefit of prefering
neutrinos with $nonzero$ momentum.
Once again the source and
detector will be modeled by harmonic oscillators.  This
time, however, the oscillators will be coupled to the usual
$V-A$ leptonic current rather than simply to the neutrino field.  
As a result, the interactions at the source and detection
points will involve both the neutrino and its associated charged
lepton.
It is convenient to take the initial state to consist only of the
source and detector, both in their first excited states.
The source decays by emitting a neutrino and its associated
charged anti-lepton, and the detector decays by 
absorbing the neutrino and emitting another charged lepton:
\begin{eqnarray}
	({\rm source})^* & \rightarrow & \nu(k)+l_{\alpha}^+(p_1)
			+ ({\rm source}) \nonumber \vspace{-.2in}\\
		& & \hspace{.15in}\hookrightarrow 
		\hspace{.15in} \nu(k) + ({\rm detector})^*
		\rightarrow l_{\beta}^-(p_2) + ({\rm detector}) .
\end{eqnarray}
This sequence of events is illustrated schematically in Fig.~\ref{spin.fg}.
The system may be described by the following
action
\equation
        S = \int d^4x \left({\cal L}_{\nu}^0 + {\cal L}_{\rm int}\right)
                + \int dt L_q^0,
\endequation
where
\begin{eqnarray}
        {\cal L}_{\nu}^0 & = &
                \sum_i\overline{\nu}_i(x)
                \left(i\dslash-m_i\right)\nu_i(x), \\
        L_q^0 & = & \sum_{\alpha}\left[\dot{q}_1^{\alpha\dagger}(t)
                \dot{q}_1^{\alpha}(t)
                - \Omega_1^2 q_1^{\alpha\dagger}(t)q_1^{\alpha}(t)
                +\dot{q}_2^{\alpha\dagger}(t)\dot{q}_2^{\alpha}(t)
                - \Omega_2^2 q_2^{\alpha\dagger}(t)q_2^{\alpha}(t)
                \right] ,\\
        {\cal L}_{\rm int} & = & -\sum_{\alpha}\left[
                \epsilon_1(t)q_1^{\alpha}(t)h_1({\bf x})
			j_{\alpha}^{0}(x) + 
                \epsilon_2(t)q_2^{\alpha}(t)h_2({\bf x})
                        j_{\alpha}^{0\dagger}(x) + {\rm h.c.}
			\right],
\end{eqnarray}
and where $j_{\alpha}^{0}(x)$ is the zeroth\footnote{In a more realistic
calculation, one might perhaps couple the $V-A$ current to a current
representing the initial and final nucleus~\cite{gkll}.  
If these nuclei are sufficiently
non-relativistic then it is a good approximation to consider only the
zeroth component of the current.} component of the
leptonic $V-A$ current
\equation
	j_{\alpha}^{\mu}(x) \equiv \sum_i{\cal U}_{\alpha i}^*
		\overline{\nu}_i\gamma^{\mu}P_Ll_{\alpha}(x),
	\;\;\;\;\;l_{\alpha}=e,\mu,\tau,\ldots,
\endequation
with $P_L$$\equiv$$(1-\gamma^5)/2$.  Once again,
$\epsilon_{1(2)}$ and $h_{1(2)}$ are functions which parametrize the 
temporal and spatial couplings of the neutrino and lepton fields to the 
source (detector).  

The calculation of the amplitude proceeds in complete analogy
with the calculation for the bosonic case and we shall omit most
of the details.  As above, we take $\epsilon_{1}(t)$ ($\epsilon_{2}(t)$)
to be a gaussian of width $\sigma_{t_1}$ ($\sigma_{t_2}$) centered
at $t$$=$$0$ ($t$$=$$t_D$) and $h_1({\bf x})$ ($h_2({\bf x})$)  
to be a gaussian of width $\sigma_{x_1}$ ($\sigma_{x_2}$) centered
at ${\bf x}$$=$$0$ (${\bf x}$$=$${\bf x}_D$) (see Eqs.~(\ref{psi1def}),
(\ref{psi2def}), (\ref{eps1def}) and (\ref{eps2g})).
Thus, we omit here the case of the (coherent) ``step function''
detector and consider only the (incoherent) ``gaussian'' detector.
Also, recall that the energies of the source and detector
are $\Omega_1$ and $\Omega_2$, respectively.
The amplitude to detect a neutrino of flavour $\beta$ given that
a neutrino of flavour $\alpha$ was emitted at the source is then
given by
\begin{eqnarray}
	{\cal A}_{\alpha\rightarrow\beta} 
		& = & (2\pi)\epsilon_1^0\epsilon_2^0\sigma_{t_1}
		\sigma_{t_2}\sum_i{\cal U}_{\beta i}{\cal U}_{\alpha i}^*
		\int\frac{d^3k}{(2\pi)^3 2E_i} \exp\left[
                -\frac{1}{2}(\Omega_1-E(p_1)-E_i)^2\sigma_{t_1}^2
			\nonumber \right.\\
		& & \left.\!\!\!
                -\frac{1}{2}(\Omega_2+E_i-E(p_2))^2\sigma_{t_2}^2
		-\frac{1}{2}|{\bf k}+{\bf p}_1|^2\sigma_{x_1}^2
                -\frac{1}{2}|{\bf k}-{\bf p}_2|^2\sigma_{x_2}^2
		-iE_it_D+i{\bf k}\cdot{\bf x}_D \right]
			\nonumber \\
		& & \!\!\! \times \;\;\overline{u}_{\beta}(p_2)\gamma^0
		P_L\left(\kslash + m_i\right)\gamma^0P_L
		v_{\alpha}(p_1) ,
	\label{adirac}
\end{eqnarray}
in which the subscripts on the $u$ and $v$ spinors refer to their
flavours; the spinors also have an implicit spin index which has
been omitted.

The above expression for the amplitude is qualitatively
similar to the analogous expression, Eq.~(\ref{agauss2}), 
derived previously in the bosonic model, with a few notable
exceptions.  On a technical note, we see first that
it is no longer possible to perform the angular parts
of the ${\bf k}$ integral exactly as was done in the previous
case.  This occurs because of the presence of the momenta
of the charged leptons,
${\bf p}_1$ and ${\bf p}_2$, which complicate the
integrand somewhat.  A related point is that now the neutrinos'
momenta are not centered around zero, as was the case above.
Rather, we have for the momenta
\begin{eqnarray}
	{\bf k} & \approx & -{\bf p}_1 \label{k1cons}, \\
	{\bf k} & \approx & {\bf p}_2 \label{k2cons} 
\end{eqnarray}
and for the energies
\begin{eqnarray}
	\Omega_1 & \approx & E(p_1) + E_i , \label{en1cons}\\
	E_i + \Omega_2 & \approx & E(p_2) , \label{en2cons}
\end{eqnarray}
where $E_i$ is the energy of the $i^{\em th}$ neutrino
mass eigenstate.
The relations (\ref{k1cons})--(\ref{en2cons}) are only
approximate equalities since the degree to which each of
them holds is determined by the relative sizes of 
$\sigma_{x_1},\ldots,\sigma_{t_2}$.  
The fact that the neutrinos' momenta are not centered
about the origin is rather encouraging because it indicates
that this model would not be expected to have the (unphysical)
feature that it favours non-relativistic neutrinos, as was
the case in the bosonic model of the previous section. 
The final difference, compared to the bosonic case, is
the presence of the matrix element,
$\overline{u}_{\beta}\ldots v_{\alpha}$, which contains
all of the information regarding the neutrinos' spins.
It is interesting to note the presence of the factor
\equation
	\frac{(\kslash +m_i)}{2E_i},
\endequation
which arises in this case in part due to the sum over spins of
the neutrino $u$ spinors, $\sum_s u^s(k_i)\overline{u}^s(k_i)$.  
This same factor
appears in the field theoretic
calculation of Ref.~\cite{gkll}, but
in that case is due to an integral in the complex $k_0$ plane
which extracts the pole of the propagator~\cite{endgrimus}.  
We need not do any such integration since we always insist 
that our source be turned ``off'' before our detector is turned ``on.''
This forces the neutrinos to always be on-shell.

It would be possible at this point to proceed as we did in the
previous section.  First we could examine the response of the detector
to the source by looking very carefully at the case in which there
is only one neutrino.  Armed with this knowledge we could define
the probability in analogy with the bosonic case and study its
behaviour as a function of the various parameters of the theory.
While this progam might be deserving of future study,
for now we shall content ourselves with a more qualitative
examination of the generic features of this model.

As we have noted, two of the main qualitative differences
between this model and our former bosonic model
are the different energy-momentum conservation equations
and the presence of the matrix element in the integrand.
A further difference is that in order to obtain the oscillation
probability, we now need to integrate over the momenta
of the two outgoing charged leptons.  Since the ${\bf k}$ integral
in the expression for the amplitude is 
expected to be dominated by values of ${\bf k}$ which are
parallel to ${\bf x}_D$~\cite{grimus}, 
the ${\bf p}_1$ and ${\bf p}_2$ integrals
would similarly be dominated by values anti-parallel and parallel,
respectively, to ${\bf x}_D$, due to the damping terms in the
exponential of Eq.~(\ref{adirac}).  In order to get some idea
of the effect of the matrix element as a function of the
neutrino's mass, then, let us evaluate it when all of the
momenta are parallel (or anti-parallel) to ${\bf x}_D$.  
(A somewhat similar analysis to the following may be found
in Ref.~\cite{noweakst}.)  Choosing an
explicit representation for the gamma matrices and 
adopting the normalization conditions of Itzykson and 
Zuber~\cite[pp. 57, 145-6, 201]{itzykson}, we find that only
two of the four helicity combinations
of the leptons survive, yielding
\begin{eqnarray}
	{\cal M}^{++}_{\alpha\rightarrow\beta}(m_i) 
		& = & -(E_i-k)\frac{(E(p_1)+m_{\alpha}+p_1)
				(E(p_2)+m_{\beta}-p_2)}
		{2\left[4m_{\alpha}m_{\beta}(E(p_1)+m_{\alpha})
				(E(p_2)+m_{\beta})\right]^{1/2}} ,\\
        {\cal M}^{--}_{\alpha\rightarrow\beta}(m_i) 
		& = & -(E_i+k)\frac{(E(p_1)+m_{\alpha}-p_1)
                                (E(p_2)+m_{\beta}+p_2)}
                {2\left[4m_{\alpha}m_{\beta}(E(p_1)+m_{\alpha})
                                (E(p_2)+m_{\beta})\right]^{1/2}} ,
\end{eqnarray}
where $k$$\equiv$$|{\bf k}|$, etc., and where the
``$++$'' and ``$--$''
superscripts refer to the helicities of the lepton and
anti-lepton.  In the limit as the neutrino mass goes to zero,
only the combination in which both leptons have negative helicity
survives, since the exchanged neutrino can only have negative 
helicity in that limit.  For non-zero masses it becomes possible to
also produce lepton pairs with positive helicity.

The quantities which will occur in the oscillation probability
are the squares of the matrix elements.  Let us define
\begin{eqnarray}
	h^{+}_{\alpha\rightarrow\beta}(m_i)
	 	& = & |{\cal M}^{++}_{\alpha\rightarrow\beta}(m_i)|^2/
			|{\cal M}^{--}_{\alpha\rightarrow\beta}(0)|^2 , \\
        h^{-}_{\alpha\rightarrow\beta}(m_i)
                & = & |{\cal M}^{--}_{\alpha\rightarrow\beta}(m_i)|^2/
                        |{\cal M}^{--}_{\alpha\rightarrow\beta}(0)|^2 .
\end{eqnarray}
Then $h^{+}$ ($h^{-}$) gives some measure of the probability
that the source/detector interaction gives rise to two leptons
with positive (negative) helicity.
Since the efficiency of the system at producing and detecting
neutrinos of a given mass is determined to some extent by the functions
$h^{\pm}$, it is useful to plot them as a function of the mass
of the exchanged neutrino.

\begin{figure}[tb]
\epsfbox[65 386 492 705]{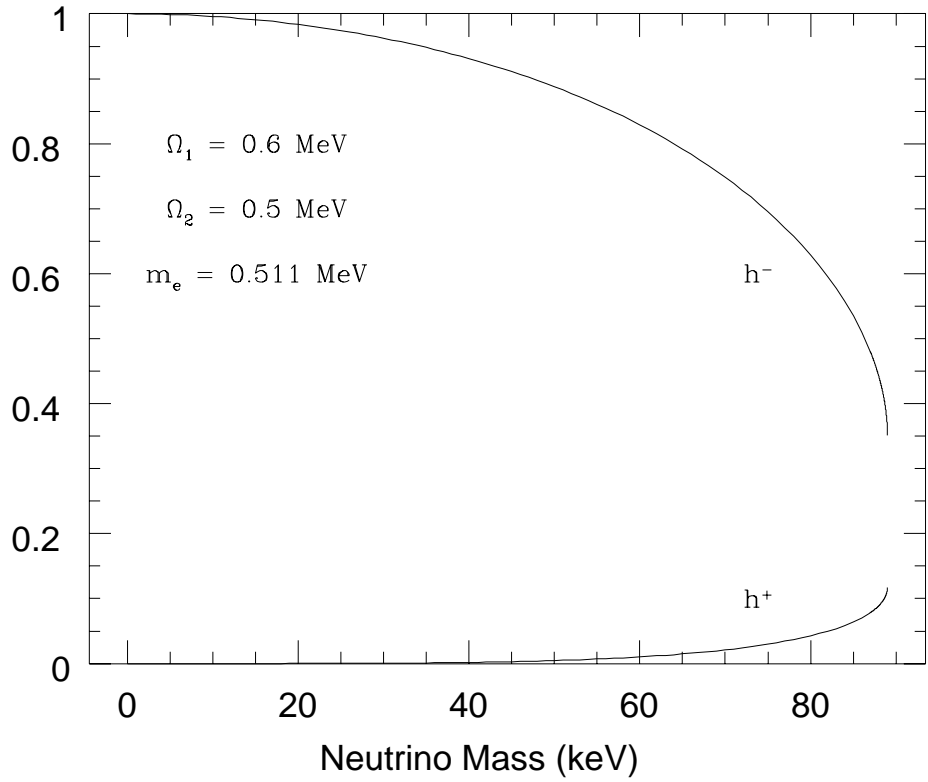}
\caption[Plot of the matrix element squared as a function of $m$]
{Plot of the two functions
$h^{+}_{e\rightarrow e}$ and $h^{-}_{e\rightarrow e}$ as a function
of the neutrino mass.  These provide a measure of the probability
to produce lepton pairs with helicity $+1$ and
$-1$, for $h^{+}$ and $h^{-}$, respectively.}
\label{matrixsq.fg}
\end{figure}

It turns out that the energy-momentum conservation equations,
Eqs.~(\ref{k1cons})--(\ref{en2cons}), are over-complete.  Thus, for 
given values of the charged lepton and neutrino masses, 
for example, $\Omega_1$
and $\Omega_2$ may be found such that all of the conditions are
met, but when the neutrino mass is varied, at least one
of the conditions needs to be violated.  This problem is
related to the difficulty which occurred in the bosonic
model (where momenta close to zero were favoured) and has
its root in the fact that our source and detector are
fixed and do not recoil.  For the purposes of our plot, let us
require that Eqs.~(\ref{k1cons}), (\ref{en1cons}) and (\ref{en2cons})
hold exactly -- so that energy and momentum are conserved at the source
and energy is conserved at the detector -- 
and allow the momentum conservation at the detector,
Eq.~(\ref{k2cons}), to be violated.
As in our previous model,
this can again be allowed by setting $\sigma_{x_2}$ to be somewhat
small.\footnote{
On physical grounds we would prefer to allow momentum conservation
to be violated somewhat rather than energy conservation.  
The reason for this is
that in the former case, the small value required for $\sigma_x$ is
still of a reasonable magnitude compared to nuclear scales
(it is on the order of several hundred
fm in the example considered here), but the value which would
be required for $\sigma_t$ would be far too small compared to any
time scales in the physical problem.}
For the plot let us take $\alpha$$=$$\beta$$=e$,
so that both the source and detector are sensitive to electron
neutrinos.  We then set
\begin{eqnarray}
		\Omega_1&=&0.6\; {\rm MeV},
			\;\;\;\Omega_2 = 0.5\; {\rm MeV}, \nonumber \\
		m_{\alpha} & = & m_{\beta} = m_e =  0.511\; {\rm MeV} .
\end{eqnarray}
Fig.~\ref{matrixsq.fg} shows a plot of $h^{+}_{e\rightarrow e}(m)$
and $h^{-}_{e\rightarrow e}(m)$ as a function of the neutrino
mass.  The ``threshold'' in this case is determined by the condition
$\Omega_1$$=$$m_e+m$, where $m$ is the neutrino mass.
The upper curve corresponds to the negative helicity case
and approaches unity as $m$$\rightarrow$$0$.  The lower 
curve disappears in the same limit.  For neutrino masses closer
to threshold, fairly substantial deviations from
the $m$$=$$0$ case are observed to occur.

The plot in Fig.~\ref{matrixsq.fg} should of course be treated
with some caution, since it shows only the square of the matrix element
evaluated at some ``optimal'' energy and momentum configuration.
In general, the oscillation probability will also receive contributions
due to energy and momentum configurations which are non-optimal.
Furthermore, it has been found that the procedure which we have followed
can lead to non-sensical results if the neutrino mass is taken to be 
large compared to the lepton mass.\footnote{This occurs because,
in our prescription, $k$ and $p_2$ need not be the same.  For very
heavy neutrinos this starts to cause problems in this approach.}
In any case, however, the plot {\em does} demonstrate something which might
be regarded as ``typical'':  for non-relativistic
neutrinos there will be a non-zero probability to produce 
charged leptons in the final state which have the ``wrong''
helicity configurations.  Thus, particularly if the spin
of the leptons were to be measured in a certain experiment, one could
expect there to be quite strong mass effects for non-relativistic
neutrinos.  In our case, for example, there is a suppression
of the negative helicity final states for large mass
and a mild enhancement of the positive helicity ones.

Since in this model the neutrinos no longer have their momenta
centered about the (unphysical) value of ``zero,'' one would expect
in this case that the non-relativistic neutrinos would not be
favoured, as was found to be the case in the bosonic model studied above.
In fact, it is possible that there would be a suppression for
non-relativistic neutrinos due to the phase space suppression
of the final state leptons, for small momenta.  This question could really
only be answered by performing a thorough numerical analysis of
the model, which we shall not do at this time.

\section{Discussion and Conclusions}
\label{sec3.4}

Most phenomenological work in the field of Particle ($\nu$, $K$, $B$, ...)
oscillations describes the oscillations as a function of time and then
converts the time dependence of the results to a space dependence.
There have been many attempts in the literature to improve on 
these calculations by explicitly including the spatial dependence
of the wave function. These approaches have necessarily lead to 
the description of the wave function as a wave packet.
It has been shown that several recent claims that such
wave packet approaches lead to different results than the
simple time--oscillation approach, are incorrect and that a
{\em proper} wave packet calculation leads to the ``expected'' results.

In this paper we have presented a novel approach to the study of the
spatial dependence of neutrino (and other particle) oscillations.
We have done this by coupling the neutrino field to an
idealized, localized model of a source and detector which
we have chosen to describe as simple harmonic oscillators which can
be excited or de-excited by the absorption or emission of a neutrino.
The system begins with the source in the first excited state and
the detector in its ground state. We then compute the probability
that at a much later time the source is in its ground state (so that
it has emitted a neutrino) and the detector is in its first excited
state (so that it has absorbed a neutrino). This probability is 
evaluated as a function of the distance between the source and the
detector and it depends, in detail, on the spatial extent of the
source and the detector as well as on the length of time for which
each is on. We have seen how to use this dependence to obtain
a better understanding of how neutrino oscillations depend on the
time resolution and the 
coherence properties of the source and the detector. 
We have also seen how our approach is useful in clarifying several
subtle issues related to the Quantum Mechanics of neutrino
oscillations.

\acknowledgements{We wish to thank H. Lipkin,
J. Oppenheim and B. Reznik for helpful conversations.  
We are also particularly indebted to W. Unruh for suggesting
the use of a source and detector to study this problem.
This work was supported in
part by the Natural Sciences and Engineering Research
Council of Canada. Their support is gratefully acknowledged.
Ken Kiers is also supported in part by the U.S. Department
of Energy under contract number DE-AC02-76CH00016.
Nathan Weiss acknowledges the support
of the Weizmann Institute of Science as well as the support of
this work by the Israel Science Foundation under grant number 255/96-1.}    

\appendix

\section{Approximate Amplitude for the Coherent Detector}
\label{appendch3.2}

In this appendix we shall derive an approximation
for the $t_2$$\rightarrow$$\infty$ limit of
the integral given in Eq.~(\ref{astep}) and investigate
under what circumstances the approximation is valid.
 
The form for the integral given in Eq.~(\ref{astep})
is convenient for numerical work, but is not particularly
convenient for the limit which we wish to consider.
Let us instead go back to the definition of this expression,
gotten by inserting Eq.~(\ref{eps2s}) into Eq.~(\ref{ampeps2}).
We may now formally take the limit as $t_2$$\rightarrow$$\infty$ 
by giving $\Omega_2$ a small imaginary piece.  This yields
\begin{eqnarray}
    {\cal A}_{\rm step}(x_D,t_1,\infty) & = & -i\tilde{N}
            \int_m^{\infty} \frac{dE}{E-\Omega_2-i\epsilon}
            \exp\left[-\frac{1}{2}(E-\Omega_1)^2\sigma_{t_1}^2
            \right. \nonumber \\
            & & \left. -\frac{1}{2}k^2(\sigma_{x_1}^2+\sigma_{x_2}^2)
            -i(E-\Omega_2)t_1 \right]\sin(kx_D),
\end{eqnarray}
where the limit $\epsilon$$\rightarrow$$0^+$ is understood.  This
integral may be simplified by employing the relation
\equation
    \frac{1}{E-\Omega_2-i\epsilon} = i\pi \delta(E-\Omega_2)
            + PP\frac{1}{E-\Omega_2}
\endequation
to obtain
\begin{eqnarray}
    {\cal A}_{\rm step}(x_D,t_1,\infty) & = & 
        \tilde{N}\pi\exp\left[-\frac{1}{2}
        (\Omega_2-\Omega_1)^2\sigma_{t_1}^2
        -\frac{1}{2}(\Omega_2^2-m^2)
        (\sigma_{x_1}^2+\sigma_{x_2}^2)\right]
        \sin(\bar{k}x_D) \nonumber \\
        & & -i\tilde{N} PP\int_m^{\infty}\frac{dE}{E-\Omega_2}
            \exp\left[-\frac{1}{2}(E-\Omega_1)^2\sigma_{t_1}^2
            \right. \nonumber \\
            & & \left.\;\; -\frac{1}{2}k^2(\sigma_{x_1}^2+\sigma_{x_2}^2)
            -i(E-\Omega_2)t_1 \right]\sin(kx_D),
        \label{astepap1}
\end{eqnarray}
where we have defined 
\equation   
    \bar{k}\equiv\sqrt{\Omega_2^2-m^2} .
\endequation            
In order to approximate Eq.~(\ref{astepap1}) it
is useful to make a change of variables.  On the interval
$(m,\Omega_2)$ we define $\tilde{E}$$=$$\Omega_2-E$ and
on $(\Omega_2,\infty)$ we define $\tilde{E}$$=$$E-\Omega_2$.
Then the integral in (\ref{astepap1}) may be approximated by
\begin{eqnarray}
    & &\hspace{-.5in} 
        i\tilde{N}\int_0^{\Omega_2-m}\frac{d\tilde{E}}{\tilde{E}}
        \left\{\exp\left[i\tilde{E}t_1
        -\frac{1}{2}(\tilde{E}-\Delta\Omega)^2\sigma_{t_1}^2 
        \right. \right. \nonumber \\
    & & \left.
        -\frac{1}{2}((\tilde{E}-\Omega_2)^2-m^2)
        (\sigma_{x_1}^2+\sigma_{x_2}^2)\right]
        \sin\left(\sqrt{(\tilde{E}-\Omega_2)^2-m^2}x_D\right)
        \nonumber \\
    & & -\exp\left[-i\tilde{E}t_1
        -\frac{1}{2}(\tilde{E}+\Delta\Omega)^2\sigma_{t_1}^2 
        \right. \nonumber \\
    & & \left. \left.
        -\frac{1}{2}((\tilde{E}+\Omega_2)^2-m^2)
        (\sigma_{x_1}^2+\sigma_{x_2}^2)\right]
        \sin\left(\sqrt{(\tilde{E}+\Omega_2)^2-m^2}x_D\right)\right\},
    \label{astepap2}
\end{eqnarray}
where $\Delta\Omega$$\equiv$$\Omega_2-\Omega_1$ and where the only
approximation so far is that the interval $(\Omega_2,\infty)$
has been truncated to $(\Omega_2,2\Omega_2-m)$.  This approximation
is valid if the major contribution to the integral comes from
energies close to $\Omega_2$.  In order to further approximate
the integral, let us make the {\em ansatz} that the integral
in (\ref{astepap2}) is dominated by values so close to 
$\tilde{E}$$=$$0$ that is valid to set $\tilde{E}$$=$$0$ in
the gaussian pieces.  At the end of the calculation we will be
able to see in which cases this is a reasonable approximation.
When dealing with the oscillating terms we must be a bit more 
careful.  Writing the sine's in terms of exponentials
and Taylor-expanding the arguments to first order in $\tilde{E}$
(which essentially amounts to ignoring the spreading of the 
wave packets)
leads to the following approximation for (\ref{astepap2})
\begin{eqnarray}
    & & \hspace{-.5in}
        \frac{1}{2}\tilde{N}\exp\left[-\frac{1}{2}
        (\Omega_2-\Omega_1)^2\sigma_{t_1}^2
        -\frac{1}{2}(\Omega_2^2-m^2)
        (\sigma_{x_1}^2+\sigma_{x_2}^2)\right]
        \nonumber \\
    & & \times
        \int_0^{\Omega_2-m}\frac{d\tilde{E}}{\tilde{E}}
        \left[e^{i\tilde{E}t_1}\left(
        e^{i(\bar{k}-\tilde{E}/\bar{v})x_D}-
        e^{-i(\bar{k}-\tilde{E}/\bar{v})x_D}\right) \right. \nonumber \\
    & & \left.\;\; -e^{-i\tilde{E}t_1}\left(
        e^{i(\bar{k}+\tilde{E}/\bar{v})x_D}-
        e^{-i(\bar{k}+\tilde{E}/\bar{v})x_D}\right) \right] \nonumber \\
    & = & -i\tilde{N}\exp\left[\ldots\right]
        \int_0^{\Omega_2-m}\frac{d\tilde{E}}{\tilde{E}}
        \left[e^{i\bar{k}x_D}\sin\left(
        \tilde{E}\left(\frac{x_D}{\bar{v}}-t_1\right)\right)
        \right. \nonumber \\
    & & \left. \;\;
        +e^{-i\bar{k}x_D}\sin\left(
        \tilde{E}\left(\frac{x_D}{\bar{v}}+t_1\right)\right)\right],
\end{eqnarray}
where
\equation   
    \bar{v}\equiv \frac{\sqrt{\Omega_2^2-m^2}}{\Omega_2} .
\endequation

The final step in the approximation is to note that, if
\equation
    \frac{x_D}{\bar{v}}\pm t_1\gg\sigma_{t_1}
	\label{timecondition}
\endequation
and if $|\Delta\Omega\sigma_{t_1}|$ is less than or of order unity,
then we may approximate the sine terms by
delta functions, since
\equation
    \lim_{L\rightarrow\infty} \frac{\sin(xL)}{x} = \pi \delta(x).
\endequation
This brings us to the desired result
\begin{eqnarray}
    {\cal A}_{\rm step}(x_D,t_1,\infty) & \simeq & 
        -i\tilde{N}\pi\exp\left[i\bar{k}x_D
        -\frac{1}{2}(\Omega_2-\Omega_1)^2\sigma_{t_1}^2
        \right. \nonumber \\
        & & \left. \;\;\;\;
        -\frac{1}{2}(\Omega_2^2-m^2)
        (\sigma_{x_1}^2+\sigma_{x_2}^2)\right].
        \label{astepap3}
\end{eqnarray}

Note that the condition in Eq.~(\ref{timecondition}) simply requires
that the detector be turned on before any appreciable amount of
flux reaches it.


\begin{thebibliography}{99}
\bibitem{endgrimus} This connection has also been noted in
        Ref.~\cite{grimus}.
\end{thebibliography}
\end{document}